\documentclass[11pt,aps,preprint,nofootinbib]{revtex4}
\usepackage[active]{srcltx}
\usepackage[utf8]{inputenc}
\usepackage{latexsym}
\usepackage{amsmath}
\usepackage{graphicx,color}

\begin{document}

\title{Supersymmetric Quantum Spherical Spins}

\author{L. G. dos Santos}
\email{lgsantos@uel.br}
\affiliation{Departamento de F\'isica, Universidade Estadual de Londrina, \\
Caixa Postal 10011, 86057-970, Londrina, PR, Brasil}

\author{L. V. T. Tavares }
\email{ltavares@uel.br}
\affiliation{Departamento de F\'isica, Universidade Estadual de Londrina, \\
Caixa Postal 10011, 86057-970, Londrina, PR, Brasil}

\author{P. F. Bienzobaz}
\email{paulabienzobaz@uel.br}
\affiliation{Departamento de F\'isica, Universidade Estadual de Londrina, \\
Caixa Postal 10011, 86057-970, Londrina, PR, Brasil}

\author{Pedro R. S. Gomes}
\email{pedrogomes@uel.br}
\affiliation{Departamento de F\'isica, Universidade Estadual de Londrina, \\
Caixa Postal 10011, 86057-970, Londrina, PR, Brasil}

%%%%%%%%%%%%%%%%%%%%%%%%%%%%%%%%%%%%%%%%%%%%%%%%%%%%%%

\begin{abstract}

In this work we investigate properties of a supersymmetric extension of the quantum spherical model from an off-shell formulation directly in the superspace. 
This is convenient to safely handle the constraint structure of the model in a way compatible with supersymmetry. 
The model is parametrized by an interaction energy, $U_{{\bf r},{\bf r}'}$, which governs the interactions between the superfields of different sites. 
We briefly discuss some consequences when $U_{{\bf r},{\bf r}'}$ corresponds to the case of first-neighbor interactions. 
After computing the partition function via saddle point method for a generic interaction, $U_{{\bf r},{\bf r}'}\equiv U(|{\bf r}-{\bf r}'|)$, we focus in the mean-field version, which reveals an interesting critical behavior. In fact, the mean-field supersymmetric model exhibits a quantum phase transition without breaking supersymmetry at zero temperature, as well as a phase transition at finite temperature with broken supersymmetry. We compute critical exponents of the usual magnetization and susceptibility in both cases of zero and finite temperature. Concerning the susceptibility, there are two regimes in the case of finite temperature characterized by distinct critical exponents. The entropy is well behaved at low temperature, vanishing as $T \rightarrow 0$.

\end{abstract}
\maketitle

%%%%%%%%%%%%%%%%%%%%%%%%%%%%%%%%%%%%%%%%%%%%%%%%%%%%%%%%%%%%%%%%%%%%%%%%%%%%%

\section{Introduction}\label{S1}

\subsection{Motivations}

It has long been noticed that supersymmetry can be realized in certain systems of condensed matter \cite{Friedan,Friedan1,Qiu}. Roughly, it can occur in systems involving both bosonic and fermionic degrees of freedom or at least in systems where the basic degrees of freedom effectively behave as bosonic and fermionic ones. Upon tuning one or more parameters of the model, supersymmetry can eventually be reached. In this case, we think of it as an emergent supersymmetry.  

One of the most interesting models where such a mechanism occurs is the tricritical Ising model in two dimensions \cite{Friedan1,Qiu}. 
To go a little deeper into this system it is convenient to consider the conformal field theory (CFT) description of statistical mechanical models \cite{Francesco,Mussardo}. Such a description relies on the fact that scale invariance, for systems with short-ranged interactions, can be extended to conformal invariance at the critical point in two dimensions. One special feature of the tricritical Ising model is that in its CFT incarnation it is represented by a superconformal theory, i.e., a field theory that, in addition to the conformal invariance, exhibits supersymmetry \cite{Friedan1,Qiu}. Thus the CFT connection unveils a hidden supersymmetry in the tricritical Ising model. 

We remind that a microscopic realization of the tricritical Ising model is given in terms of the Blume-Emery-Griffiths model \cite{Blume}, which involves a spin-1 variable, $S_i=0,\pm1$. It can be also formulated in terms of two types of variables, a spin-1/2 variable, $\sigma_i=\pm1$, and the vacancy $t_i=0,1$. They are connected through $S_i\equiv t_i\sigma_i$. This model has a rich structure since it has more than one order parameter, as $\langle S_i \rangle$ and $\langle S_i^2 \rangle$, therefore opening the possibility for exhibiting tricritical behavior and has been applied in the description of the lambda transition in mixtures of $^3$He and $^4$He \cite{Blume}. The variables $\sigma_i$ and $t_i$ can be naively  thought as fermionic and bosonic counterparts such that for certain values of the involved parameters (corresponding to the tricritical point) the model effectively behaves in a supersymmetric way which is captured in the CFT description.  Models like this provide an interesting interplay between supersymmetry and phase transitions. 

Earliest studies of supersymmetry in spin models can be found in \cite{Nicolai1,Nicolai2,Rittenberg}.
More recently, there has been much interest in identifying models with potential to exhibit supersymmetric behavior as well as the basic ingredients. In this context, 
supersymmetry has been reported to emerge in certain lattice models \cite{Fendley,Lee,Affleck} and also in topologically ordered systems \cite{Ponte,Grover,Yao,Jian}. 
These studies are appealing as they bring together ingredients of great interest as the supersymmetry, supersymmetry breaking and the quantum critical behavior.
Pursuing these lines, in this work we propose to investigate the properties of a supersymmetric extension of the so-called quantum spherical model, which is a theoretical model amenable to a number of exact calculations.

The quantum spherical model is the quantized version of the classical spherical model introduced many years ago by Berlin and Kac \cite{Berlin}. It belongs to a rare class of models, which are exactly soluble in arbitrary dimensions even in the presence of an external field. Furthermore, for hypercubic lattices in $2<d<4$ it exhibits a nontrivial critical behavior. For these reasons, along with the Ising model, the spherical model constitutes an excellent prototype to investigate properties of the critical behavior \cite{Joyce}.

Quantized versions of the spherical model go back to \cite{Obermair,Henkel} and in more recent years to \cite{Nieuwenhuizen,Vojta1,Gracia}. 
In \cite{Nieuwenhuizen} the introduction of quantum fluctuations was proposed as natural mechanism to fix the anomalous low-temperature behavior of the classical counterpart (the entropy diverges for $T\rightarrow 0$). It is known, in turn, that  quantum fluctuation due to the Heisenberg's uncertainty relation can drive a phase transition at zero temperature \cite{Thomas,Matthias,Sachdev}. In this context, it was shown in \cite{Vojta1} that in addition to the finite-temperature critical behavior the quantum spherical model exhibits a quantum phase transition, i.e., a phase transition at zero temperature. Both classical and quantum versions of the spherical model have interesting correspondence to the large-$N$ limit of classical Heisenberg model \cite{Stanley} and nonlinear sigma model \cite{Vojta1,Gomes2}, respectively. 

In view of this, the supersymmetric extension of the quantum spherical model immediately places the supersymmetry in a rich context where both thermal and quantum fluctuations may drive a phase transition. At the same time, it is an opportunity to explore a relatively simple model containing a number of interesting and nontrivial properties, which are shared with other systems of great interest. We know that supersymmetry is broken by the temperature, essentially due to the distinct thermal distributions for bosons and fermions \cite{Girardello,Daniel}. Thus, any finite-temperature phase transition occurs with broken supersymmetry. However, at zero temperature, a quantum phase transition may or may not  involve a spontaneous breaking of supersymmetry. We shall investigate these questions in this work. 

Contrarily to the models where supersymmetry is reached after some fine-tuning of the involved parameters, in our construction supersymmetry is not emergent. Indeed, it is used as the starting point to construct the model as a required symmetry. Nevertheless, the resulting model is better thought as describing effective quantum degrees of freedom. In addition to the usual spherical spin variable $S_{\bf r}$ attached to each site, where $-\infty<S_{\bf r}<\infty$ are subjected to $\sum_{\bf r}S_{\bf r}^2=N$, we consider the fermionic counterparts $\psi_{\bf r}$ and $\bar{\psi}_{\bf r}$ (actually we need also an auxiliary bosonic degree of freedom $F_{\bf r}$ in order to obtain an off-shell supersymmetry). The supersymmetric model is exactly soluble in arbitrary dimensions and, as we shall discuss, possesses interesting critical behavior. 

%%%%%%%%%%%%%%%%%%%%%%%%%%%%%%%%%%%%%%%%%%%%%%%%%%%%%%%%%%%%%%%%%%%%%
\subsection{Comparison with Previous Works and Main Results}

The first proposal of a supersymmetric extension of the quantum spherical model was presented in the reference \cite{Gomes}. There, a supersymmetric version was obtained by starting with an on-shell formulation. One subtle point, specially when we are in an on-shell description, is to conciliate the supersymmetry requirements with the constraint structure of the spherical model. This has not been handled in a fully satisfactory way in \cite{Gomes}\footnote{We will point this out precisely in   Sec. \ref{S3}}. Thus, although the treatment done in \cite{Gomes} does not affect the general pattern of critical behavior at finite temperatures, it led to a weird prediction for the critical behavior at zero temperature in the supersymmetric regime, namely, that the model does not exhibit quantum phase transition when supersymmetry is not broken. We evade such difficulties in the present work by starting with an off-shell formulation directly in the superspace, where the supersymmetry is manifest. 

The resulting model coincides with that one constructed in \cite{Bienzobaz} in the context of stochastic quantization, by exploring the mapping between a $d$-dimensional field theory and a $(d+1)$-dimensional one, when the fictitious time is not eliminated. As it is known \cite{Parisi}, this yields to a supersymmetry in the fictitious time direction. It is remarkable that the stochastic quantization prescription automatically keeps on the track all the subtleties involving the supersymmetry and the constraints. What is behind this is that the Langevin equation constitutes an explicit realization of the so-called Nicolai map \cite{Nicolai}, that is useful in the characterization of supersymmetric theories via functional integration measures. In addition, in \cite{Bienzobaz}, it was determined the critical dimensions for an interaction whose Fourier transform is parametrized by $\widehat{U}({\bf q})\sim |{\bf q}|^{\frac{x}{2}}$.

All the interactions between the bosonic and fermionic variables of different sites are given in terms of only one interaction energy, $U_{{\bf r},{\bf r}'}$. This is a requirement of supersymmetry once independent interactions in the model would lead to an explicit breaking of supersymmetry. Interesting physical properties can be extracted by simply considering the on-shell formulation, where we eliminate the auxiliary degrees of freedom. In particular, we show that competing interactions in the bosonic sector can arise even when $U_{{\bf r},{\bf r}'}$ involves only first-neighbor interactions. We illustrate this point in the case of a two-dimensional square lattice, but the conclusion extends to higher-dimensional lattices. Therefore the supersymmetric model has the potential to exhibit a Lifshitz point \cite{Hornreich,Frachebourg}.

We also present a detailed study of the mean-field version of the supersymmetric spherical model, where all the expressions are made rather explicit. 
Although this version is not able to capture the Lifshitz point, it exhibits an interesting critical behavior unveiling certain general features of the phases of the model and the supersymmetry breaking pattern. There are phase transitions governed by the thermal fluctuations at finite temperature as well as a phase transition governed by the quantum fluctuations at zero temperature. In general, the supersymmetry is broken at finite temperature due to distinct bosonic and fermionic thermal distributions. In this situation two regimes emerge in our model leading to different critical exponents for the susceptibility. They correspond to different saddle point values of the involved parameters, which happens only in the case of finite temperature. At zero temperature the only possible solution is the one in which the bosonic and fermionic frequencies are the same yielding to a vanishing ground state energy. Then the model undergoes a quantum phase transition without breaking supersymmetry.  By studying the behavior of the magnetization we show that all the phase transitions are of order-disorder type.

The work is organized as follows. In Sec. \ref{S2}, we construct the supersymmetric extension of the quantum spherical model directly in the superspace.
In Sec. \ref{S3}, the on-shell formulation is considered  in order to make clearer the type of interactions that are present in the model.
The computation of the partition function is presented in Sec. \ref{S4}, where we also discuss the saddle point solutions and the supersymmetry breaking.
In Sec. \ref{S5}, the mean-field version is considered and we study the critical behavior at both zero and finite temperature.
A summary and additional comments are presented in Sec. \ref{S6}.  

%%%%%%%%%%%%%%%%%%%%%%%%%%%%%%%%%%%%%%%%%%%%%%%%%%%%%%%%%%%%%%%%%%%%%%%%%%%%%
\section{Supersymmetric Quantum Spherical Model}\label{S2}

The starting point is the quantum spherical model, which can be constructed from the classical model by introducing a kinetic term for the spherical spins, 
\begin{equation}
L= \frac{1}{2g} \sum_{\bf r} \left(\frac{d S_{\bf r}}{d t}\right)^2-\frac{1}{2}\sum_{{\bf r},{\bf r}'}J_{{{\bf r},{\bf r}'}}S_{\bf r}S_{{\bf r}'},
\label{2.1}
\end{equation}
subject to the spherical constraint 
\begin{equation}
\sum_{\bf r}S_{\bf r}^2=N.
\label{2.2}
\end{equation} 
The quantum theory is then obtained from the partition function that, in the presence of an external magnetic field $H$, reads
\begin{equation}
Z(H)= \int \mathcal{D}S \delta\left( \sum_{\bf r}S_{\bf r}^2-N \right) \exp \left[- \int_{0}^{\beta} d\tau L_{E}+H\sum_{\bf r} S_{\bf r}\right],
\label{2.3}
\end{equation}
where $L_{E}$ is the Euclidean version of (\ref{2.1}) and the integration measure stands for $\mathcal{D}S\equiv \prod_{\bf r}\mathcal{D}S_{\bf r}$.

Now we proceed with the supersymmetric generalization of the quantum spherical model. The basic idea is to introduce a fermionic partner for each bosonic variable $S_{\bf r}$, in such a way that supersymmetry is possible. In addition, we need to be careful in order to make the supersymmetry requirements compatible with the constraint structure of the spherical model. A safe way to accomplish this is by proceeding with our construction directly in the superspace, where the supersymmetry is manifest. 

\begin{figure}[!h]
\centering
\includegraphics[scale=0.7]{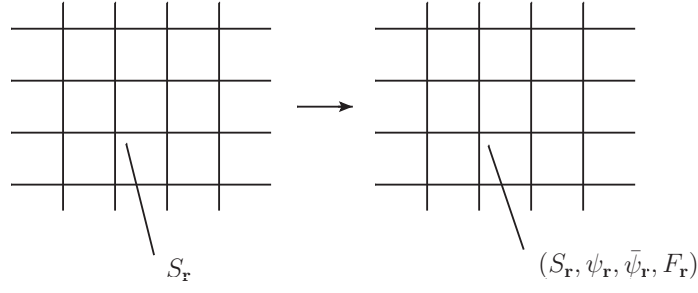}
\caption{The lattice in the left-hand-side corresponds to the case of the ordinary QSM, where there is only one spin variable $S_{\bf r}$ attached to each site. The right-hand-side corresponds to the supersymmetric extension, where there are four variables attached to each site. They can be written as components of one supervariable $\Phi_{\bf r}=S_{\bf r}+\bar{\theta}\psi_{\bf r}+\bar{\psi}_{\bf r}\theta+\bar{\theta}\theta F_{\bf r}$ . }
\label{Fig1}
\end{figure}

Quadratic spin interactions of the type $J_{{{\bf r},{\bf r}'}}S_{\bf r}S_{{\bf r}'}$ can be straightforwardly constructed from the superspace formalism if we consider the case of extended supersymmetry $\mathcal{N}=2$. In this case, the superspace consists of time, $t$, and a pair of Grassmann variables, $\theta$ and $\bar{\theta}$, which can be considered as complex conjugate of each other. The spin variable $S_{\bf r}$ gives place to a superspin variable, $\Phi_{\bf r}$, usually called a superfield. The superfield can be expanded in powers of $\theta$ and $\bar{\theta}$,
\begin{equation}
\Phi_{\bf r}=S_{\bf r}+\bar{\theta}\psi_{\bf r}+\bar{\psi}_{\bf r}\theta+\bar{\theta}\theta F_{\bf r}.
\label{2.4}
\end{equation} 
In the superfield expansion, the usual spherical spin variable $S_{\bf r}$ appears as its first component. The Grassmann variables $\psi_{\bf r}$ and $\bar{\psi}_{\bf r}$ are the fermionic counterpart, while $F_{\bf r}$ is an auxiliary bosonic degree of freedom inherent to the realization of the off-shell supersymmetry. After the theory is consistently constructed it can be integrated out, realizing thus the on-shell supersymmetry. We shall discuss this point soon. In sum, in the supersymmetric case we have more degrees of freedom per site as compared to the ordinary quantum spherical model. This is illustrated in Fig. \ref{Fig1}.

The next step is the generalization of the spherical constraint to the supersymmetric case, which is now imposed on the superfield $\Phi_{\bf r}$,
\begin{equation}
\sum_{\bf r}\Phi_{\bf r}^2=N.
\label{2.5}
\end{equation}
By comparing the corresponding powers of the Grassmann variables $\theta$ and $\bar{\theta}$ in both sides we see that (\ref{2.5}) implies, 
\begin{equation}
\sum_{\bf r}S_{\bf r}^2=N,~~~\sum_{\textbf{r}}S_{\textbf{r}}\psi_{\textbf{r}} =0,~~~\sum_{\textbf{r}}S_{\textbf{r}}\bar{\psi}_{\textbf{r}} =0,~~~\text{and}~~~\sum_{\textbf{r}}S_{\textbf{r}}F_{\textbf{r}}=\sum_{\bf r}\bar{\psi}_{\textbf{r}}\psi_{\textbf{r}}.
\label{2.6}
\end{equation}
In addition to the usual spherical constraint, compliance with supersymmetry requires a more general constraint structure. 

Supersymmetry transformations with $\mathcal{N}=2$ are generated by two supercharges, $Q$ and $\bar{Q}$, which we define as
\begin{equation}
Q\equiv -\frac{\partial}{\partial\bar\theta}-i\,\theta\frac{\partial}{\partial t}~~~
\text{and}~~~\bar{Q}\equiv \frac{\partial}{\partial\theta}+i\,\bar\theta\frac{\partial}{\partial t},
\label{2.7}
\end{equation}
satisfying the usual anticommutation relations
\begin{equation}
\{Q,Q\}=0,~~~\{\bar{Q},\bar{Q}\}=0~~~\text{and}~~~\{Q,\bar{Q}\}=-2i\frac{\partial}{\partial t}.
\label{2.8}
\end{equation}
The supercharges in (\ref{2.7}) generate translations in the superspace,
\begin{eqnarray}
&&\epsilon:~~~t\rightarrow t+i\bar\theta\epsilon,~~~\theta\rightarrow\theta-\epsilon,~~~\text{and}~~~\bar{\theta}\rightarrow\bar{\theta}\nonumber\\
&&\bar{\epsilon}:~~~t\rightarrow t-i\bar\epsilon\theta,~~~\theta\rightarrow \theta,~~~\text{and}~~~\bar\theta\rightarrow\bar\theta-\bar\epsilon,
\label{2.9}
\end{eqnarray}
where $\epsilon$ and $\bar{\epsilon}$ are infinitesimal Grassmannian parameters of the transformations. Under translations, the transformation law for the scalar superfield is $\Phi_{\bf r}'(t',\theta',\bar{\theta}')=\Phi_{\bf r}(t,\theta,\bar{\theta})$. Thus the functional variation of the superfield, defined through
$\delta\Phi_{\bf r}\equiv \Phi_{\bf r}'(t,\theta,\bar{\theta})-\Phi_{\bf r}(t,\theta,\bar{\theta})$, yields
\begin{equation}
\delta_{\epsilon}\Phi_{\bf r}=-\bar{Q}\epsilon\Phi_{\bf r}~~~\text{and}~~~\delta_{\bar{\epsilon}}\Phi_{\bf r}=-\bar{\epsilon}Q\Phi_{\bf r}.
\label{2.10}
\end{equation}
Comparison of the corresponding powers in $\theta$ and $\bar{\theta}$ furnishes the supersymmetry transformations for the components
\begin{equation}
\epsilon:~~~\delta_{\epsilon}S_{\bf r}=\bar{\psi}_{\bf r}\epsilon,~~~
\delta_{\epsilon}\psi_{\bf r}=-i\dot{S}_{\bf r}\epsilon+
F_{\bf r}\epsilon,~~~
\delta_{\epsilon}\bar{\psi}_{\bf r}=0,~~~\text{and}~~~\delta_{\epsilon}F_{\bf r}=i\dot{\bar\psi}_{\bf r}\epsilon;
\label{2.11}
\end{equation}
and
\begin{equation}
\bar{\epsilon}:~~~\delta_{\bar\epsilon}S_{\bf r}=\bar\epsilon{\psi}_{\bf r},~~~
\delta_{\bar\epsilon}\psi_{\bf r}=0,~~~
\delta_{\bar\epsilon}\bar{\psi}_{\bf r}=i\dot{S}_{\bf r}\bar\epsilon+
F_{\bf r}\bar\epsilon,~~~\text{and}~~~\delta_{\epsilon}F_{\bf r}=-i\bar\epsilon\dot{\psi}_{\bf r}.
\label{2.12}
\end{equation}

The last ingredients we need are the operators that generalize the notion of time derivative to the superspace. To this end, we introduce the supercovariant derivatives, 
\begin{equation}
D\equiv -\frac{\partial}{\partial\bar\theta}+i\theta\frac{\partial}{\partial t}~~~
\text{and}~~~\bar{D}\equiv \frac{\partial}{\partial\theta}-i\bar\theta\frac{\partial}{\partial t},
\label{2.13}
\end{equation}
which are chosen to satisfy the following anticommutation relations with the supercharges,
\begin{equation}
\{D,Q\}=\{D,\bar{Q}\}=\{\bar{D},Q\}=\{\bar{D},\bar{Q}\}=0~~~\text{and}~~~\{D,\bar{D}\}=2i\frac{\partial}{\partial t}.
\label{2.14}
\end{equation}
These relations are important since they guarantee that the supercovariant derivative of a superfield transforms as the superfield itself under supersymmetry. 
For example, by considering $D\Phi_{\bf r}$, its supersymmetry transformation is $\delta_{\epsilon}(D\Phi_{\bf r})=-\bar{Q}\epsilon (D\Phi_{\bf r})$. 
In conclusion, any action written in the superspace and involving only superfields and supercovariant derivatives of superfields, 
\begin{equation}
S=\int dt d\theta d\bar{\theta}L(\Phi_{\bf r},D\Phi_{\bf r},\bar{D}\Phi_{\bf r}),
\label{2.15}
\end{equation}
is manifestly supersymmetric. With this, we can immediately generalize the quantum spherical model (\ref{2.1}) to the supersymmetric case as
\begin{equation}
S=\int dt d\theta d\bar{\theta} \left(\frac{1}{2}\sum_{\bf r} \bar{D}\Phi_{\bf r} D\Phi_{\bf r} 
+\frac{1}{2}\sum_{{\bf r},{\bf r}'}U_{{\bf r},{\bf r}'}\Phi_{\bf r}\Phi_{{\bf r}'} \right),
\label{2.16}
\end{equation}
subject to the constraint (\ref{2.5}). The constraint can be implemented directly in the superspace action via a super Lagrange multiplier 
\begin{equation}
\Xi(t,\theta,\bar{\theta})= \gamma+\bar{\theta}\xi+\bar{\xi}\theta+\bar{\theta}\theta \mu,
\label{2.17}
\end{equation}
according to
\begin{equation}
S=\int dt d\theta d\bar{\theta} \left(\frac{1}{2}\sum_{\bf r} \bar{D}\Phi_{\bf r} D\Phi_{\bf r} 
+\frac{1}{2}\sum_{{\bf r},{\bf r}'}U_{{\bf r},{\bf r}'}\Phi_{\bf r}\Phi_{{\bf r}'} 
-\Xi\left(\sum_{\bf r}\Phi_{\bf r}^2-N\right)\right).
\label{2.18}
\end{equation}
From this action we can obtain the supersymmetric Lagrangian in terms of the components,  
\begin{eqnarray}
L_{SUSY}& =&\frac{1}{2}\sum_{\bf r}\dot{S}_{\bf r}^2
+\frac{1}{2}\sum_{\bf r}F_{\bf r}^2+i\sum_{\bf r}\bar{\psi}_{\bf r}\dot{\psi_{\bf r}}
+\sum_{{\bf r},{\bf r}'}U_{{\bf r},{\bf r}'}\left(
S_{\bf r}F_{{\bf r}'} -\bar{\psi}_{\bf r}\psi_{{\bf r}'}\right)\nonumber\\
&+&\gamma\sum_{\bf r}\left(F_{\bf r}S_{\bf r}-\bar{\psi}_{\bf r}\psi_{\bf r}\right)
-\sum_{\bf r}\bar{\psi}_{\bf r}\xi S_{\bf r}
-\sum_{\bf r}\bar{\xi}\psi_{\bf r} S_{\bf r}
-\mu\left(\sum_{\bf r}S_{\bf r}^2-N\right),
\label{2.19}
\end{eqnarray}
up to redefinitions of the Lagrange multipliers to absorb unimportant numerical factors. 
Some comments are in order. Firstly, the parameter $g$ present in (\ref{2.1}) that measures the quantum fluctuations in the system, 
must also be present in the supersymmetric model.  It can be introduced simply by a rescaling of time coordinate as $t\rightarrow \sqrt{g} t$. 
Second, as in the ordinary quantum spherical model, the interaction $U_{{\bf r},{\bf r}'}$ is assumed to be a function only on the distance between 
the sites, i.e., $U_{{\bf r},{\bf r}'}\equiv U(|{\bf r}-{\bf r}'|)$. In the next section we will discuss some physical consequences of considering explicitly the case of first-neighbor interactions between superfields.

%%%%%%%%%%%%%%%%%%%%%%%%%%%%%%%%%%%%%%%%%%%%%%%%%
\section{On-shell Formulation}\label{S3}

To make transparent the nature of the interactions involved in the supersymmetric extension of the quantum spherical model it is instructive 
to consider the on-shell formulation, which is obtained by  integrating out the auxiliary bosonic degree of freedom $F_{\bf r}$. To this end, we select the $F_{\bf r}$-dependent part of the Lagrangian in (\ref{2.19}),
\begin{equation}
L_F= \frac{1}{2}\sum_{\mathbf{r}}F_{\mathbf{r}}^{2}
+\sum_{\mathbf{r},\mathbf{r}'}U_{\mathbf{r},\mathbf{r}'}F_{\mathbf{r}}S_{\mathbf{r}'}
+\gamma\sum_{\mathbf{r}}F_{\mathbf{r}}S_{\mathbf{r}}.
\label{3.1}
\end{equation}
As it is an auxiliary field, its equation of motion is simply an algebraic one, 
\begin{equation}
\frac{\partial L_F}{\partial F_{\bf r}}=0~~~\Rightarrow~~~ F_{\textbf{r}} 
= -\gamma S_{\textbf{r}}-\sum_{\textbf{r}'}U_{\textbf{r},\textbf{r}'}S_{\textbf{r}'}.
\label{3.2}
\end{equation}
Plugging this back into the Lagrangian (\ref{3.1}), it follows
\begin{eqnarray}
L_{F} = -\frac{1}{2}\sum_{\textbf{r},\textbf{r}'}\left(\sum_{\textbf{r}''}U_{\textbf{r},\textbf{r}''}U_{\textbf{r}'',\textbf{r}'}\right)S_{\textbf{r}}S_{\textbf{r}'}
-\gamma\sum_{\textbf{r},\textbf{r}'}U_{\textbf{r},\textbf{r}'}S_{\textbf{r}}S_{\textbf{r}'}
-\frac{1}{2}\gamma^{2}N.
\label{3.3}
\end{eqnarray}
We see that there are two types of interactions between the bosonic spin variables $S_{\bf r}$, namely, $\gamma U_{\textbf{r},\textbf{r}'}$ and $J_{\textbf{r},\textbf{r}'}\equiv\sum_{\textbf{r}''}U_{\textbf{r},\textbf{r}''}U_{\textbf{r}'',\textbf{r}'}$. Putting together all the contributions, the complete on-shell Lagrangian that follows from (\ref{2.19}) is
\begin{eqnarray}
L & = & \frac{1}{2}\sum_{\mathbf{r}}\dot{S}_{\mathbf{r}}^{2}
+i\sum_{\mathbf{\mathbf{r}}}\bar{\psi}_{\mathbf{r}}\dot{\psi}_{\mathbf{r}}
-\frac{1}{2}\sum_{\textbf{r},\textbf{r}'}J_{\textbf{r},\textbf{r}'}S_{\textbf{r}}S_{\textbf{r}'}
-\sum_{\mathbf{r},\mathbf{r}'}U_{\mathbf{r},\mathbf{r}'}\bar{\psi}_{\mathbf{r}}\psi_{\mathbf{r}'}
- \sum_{\mathbf{r}}S_{\mathbf{r}}\left(\bar{\psi}_{\mathbf{r}}\xi+\bar{\xi}\psi_{\mathbf{r}}\right)\nonumber \\
 & -&\mu\sum_{\mathbf{r}}\left(S_{\mathbf{r}}^{2}-N\right)-\frac{1}{2}\gamma^{2}N
 -\gamma\left(\sum_{\textbf{r},\textbf{r}'}U_{\textbf{r},\textbf{r}'}S_{\textbf{r}}S_{\textbf{r}'}
 +\sum_{\mathbf{r}}\bar{\psi}_{\mathbf{r}}\psi_{\mathbf{r}}\right).
 \label{3.4}
\end{eqnarray}
The corresponding supersymmetry transformations are,
{\begin{equation}
\epsilon:~~~\delta_{\epsilon}S_{\bf r}=\bar{\psi}_{\bf r}\epsilon,~~~
\delta_{\epsilon}\psi_{\bf r}=-i\dot{S}_{\bf r}\epsilon
-\left(\gamma S_{\textbf{r}}+\sum_{\textbf{r}'}U_{\textbf{r},\textbf{r}'}S_{\textbf{r}'}\right)\epsilon,~~~\text{and}~~~
\delta_{\epsilon}\bar{\psi}_{\bf r}=0;
\label{3.5}
\end{equation}
and
\begin{equation}
\bar{\epsilon}:~~~\delta_{\bar\epsilon}S_{\bf r}=\bar\epsilon{\psi}_{\bf r},~~~
\delta_{\bar\epsilon}\psi_{\bf r}=0,~~~\text{and}~~~
\delta_{\bar\epsilon}\bar{\psi}_{\bf r}=i\dot{S}_{\bf r}\bar\epsilon
-\left(\gamma S_{\textbf{r}}+\sum_{\textbf{r}'}U_{\textbf{r},\textbf{r}'}S_{\textbf{r}'}\right)\bar\epsilon.
\label{3.6}
\end{equation}

It is interesting to observe the following point in the constraint structure in (\ref{3.4}). We have the same three constraints 
as in the off-shell formulation, implemented by the Lagrange multipliers $\mu$, $\xi$, and $\bar{\xi}$. On the other hand, the Lagrange multiplier $\gamma$, which in the off-shell formulation implemented the last constraint of (\ref{2.6}), in the on-shell expression (\ref{3.4}) it can be thought as implementing a constraint as an average with a Gaussian distribution instead of a delta due to the term proportional to $\gamma^2$. Its equation of motion is
\begin{equation}
\gamma=-\frac{1}{N}\left(\sum_{\textbf{r},\textbf{r}'}U_{\textbf{r},\textbf{r}'}S_{\textbf{r}}S_{\textbf{r}'}+\sum_{\mathbf{r}}\bar{\psi}_{\mathbf{r}}\psi_{\mathbf{r}}\right).
\label{3.7}
\end{equation}
By using this relation in (\ref{3.4}), we end up with 
\begin{eqnarray}
L& = & \frac{1}{2}\sum_{\mathbf{r}}\dot{S}_{\mathbf{r}}^{2}
+i\sum_{\mathbf{\mathbf{r}}}\bar{\psi}_{\mathbf{r}}\dot{\psi}_{\mathbf{r}}-\frac{1}{2}\sum_{\textbf{r},
\textbf{r}'}J_{{\bf r},{\bf r}'}S_{\textbf{r}}S_{\textbf{r}'}
-\sum_{\mathbf{r},\mathbf{r}'}U_{\mathbf{r},
\mathbf{r}'}\bar{\psi}_{\mathbf{r}}\psi_{\mathbf{r}'}
-\sum_{\mathbf{r}}S_{\mathbf{r}}\left(\bar{\psi}_{\mathbf{r}}\xi
+\bar{\xi}\psi_{\mathbf{r}}\right)\nonumber \\
 & - & \mu\sum_{\mathbf{r}}\left(S_{\mathbf{r}}^{2}-N\right)+\frac{1}{2N}\left(\sum_{\textbf{r},\textbf{r}'}U_{\textbf{r},\textbf{r}'}S_{\textbf{r}}S_{\textbf{r}'}
 +\sum_{\mathbf{r}}\bar{\psi}_{\mathbf{r}}\psi_{\mathbf{r}}\right)^{2}.
 \label{3.8}
\end{eqnarray}
Therefore, the final effect of eliminating $\gamma$ is to introduce quartic interactions between the physical variables. In the computation of the partition function in Sec. \ref{S4}, however, we will not proceed in this way. Instead, we will keep all the Lagrange multipliers and then look for a saddle point solution for $(\mu,\xi,\bar{\xi},\gamma)$, which turns out to be exact in the thermodynamic limit.

Before proceeding, it is opportune to contrast this approach with that one of Ref. \cite{Gomes}. The crucial difference is that in \cite{Gomes}, the equation of motion of the auxiliary field $F$ was used in a way independent of the constraint structure. Thus, if we compare (\ref{3.2}) with the corresponding expression in \cite{Gomes}, we see that in the later there is no contribution proportional to the Lagrange multiplier $\gamma$. The consequence is that, in the on-shell Lagrangian considered in \cite{Gomes}, the quartic terms above are absent. As already mentioned, this leads to a weird prediction for the critical behavior at zero temperature.

%%%%%%%%%%%%%%%%%%%%%%%%%%%%%%%%%%%%%%%%%%%%%%%%%%%%%%%%%%%%
\subsection{First-Neighbor Interactions}

At this point it is interesting to go back to the on-shell Lagrangian (\ref{3.4}) and see explicitly the effect of a first-neighbor interaction between the superfield variables (remember the interaction term in (\ref{2.18})), i.e., the effect of assuming the following form for $U_{\textbf{r},\textbf{r}'}$,
\begin{eqnarray}
U_{\textbf{r},\textbf{r}'} \equiv U\sum_{I=1}^{d}\left(\delta_{\textbf{r},\textbf{r}'+\mathbf{e}^{I}}+\delta_{\textbf{r},\mathbf{r}'-\mathbf{e}^{I}}\right),
\label{3.9}
\end{eqnarray}
where $U$ is the interaction energy that can be positive (ferro) or negative (anti-ferro). We are considering a $d$-dimensional hypercubic lattice with $\mathbf{e}^{I}$ being a set of orthogonal unit vectors along all directions,
\begin{eqnarray}
\left\{ \mathbf{e}^{I}\right\}  & = & \left\{ \left(1,0,\ldots,0\right);\left(0,1,0,\ldots,0\right);\ldots;\left(0,\ldots,0,1\right)\right\}.
\label{3.10}
\end{eqnarray}
Given the interaction (\ref{3.9}), we obtain for $J_{\textbf{r},\textbf{r}'}\equiv\sum_{\textbf{r}''}U_{\textbf{r},\textbf{r}''}U_{\textbf{r}'',\textbf{r}'}$,  
\begin{equation}
J_{\textbf{r},\textbf{r}'}=U^2\sum_{I,J=1}^d\left(\delta_{{\bf r},{\bf r}'+{\bf e}^I+{\bf e}^J}+\delta_{{\bf r},{\bf r}'+{\bf e}^I-{\bf e}^J}+\delta_{{\bf r},{\bf r}'-{\bf e}^I+{\bf e}^J}+\delta_{{\bf r},{\bf r}'-{\bf e}^I-{\bf e}^J}\right).
\label{3.11}
\end{equation}
We see that this expression contains interactions between second-neighbors as well as between diagonal neighbors. This is illustrated in Fig. \ref{Fig2} for the case of a two-dimensional square lattice. 
\begin{figure}[!h]
\centering
\includegraphics[scale=0.7]{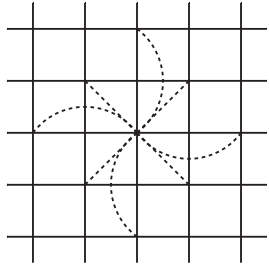}
\caption{Interactions present in (\ref{3.11}) for a two-dimensional square lattice.}
\label{Fig2}
\end{figure}

Collecting all terms involving interactions in (\ref{3.4}), we have
\begin{eqnarray}
L_{int}&\equiv&- \gamma \sum_{\textbf{r},\textbf{r}'}U_{\textbf{r},\textbf{r}'}S_{\textbf{r}}S_{\textbf{r}'}
-\frac{1}{2}\sum_{\textbf{r},\textbf{r}'}J_{\textbf{r},\textbf{r}'}S_{\textbf{r}}S_{\textbf{r}'}
-\sum_{\mathbf{r},\mathbf{r}'}U_{\mathbf{r},\mathbf{r}'}\bar{\psi}_{\mathbf{r}}\psi_{\mathbf{r}'}\nonumber\\
&=&-\gamma U\sum_{\bf r}\sum_{I=1}^d \left(S_{\bf r}S_{{\bf r}+{\bf e}^I}
+S_{\bf r}S_{{\bf r}-{\bf e}^I} \right)\nonumber\\&-&\frac{U^2}{2}\sum_{\bf r}\sum_{I,J=1}^d
\left(S_{\bf r}S_{{\bf r}+{\bf e}^I+{\bf e}^J}+S_{\bf r}S_{{\bf r}+{\bf e}^I-{\bf e}^J}+S_{\bf r}S_{{\bf r}-{\bf e}^I
+{\bf e}^J}+S_{\bf r}S_{{\bf r}-{\bf e}^I-{\bf e}^J} \right)\nonumber\\
&-&U\sum_{\bf r}\sum_{I=1}^d \left(\bar{\psi}_{\bf r}\psi_{{\bf r}+{\bf e}^I}+\bar{\psi}_{\bf r}\psi_{{\bf r}-{\bf e}^I} \right).
\label{3.12}
\end{eqnarray}
We can extract some interesting physical properties from this expression. Notice that, in addition to the first-neighbor interactions, there are also second and diagonal interactions involving the bosonic variables $S_{\bf r}$. Therefore, if the sign of $\gamma U$ is negative, this will generate a competition in the bosonic sector of the model, since $U^2$ is always positive. This type of ingredient usually produces a rich phase diagram with the presence of modulated phases and a Lifshitz point \cite{Hornreich}. 
Competing interactions were investigated in the usual quantum spherical model \cite{Frachebourg,Paula_Salinas,Nussinov,Henkel1} and it would be interesting to compare modulated phases in the nonsupersymmetric and supersymmetric situations. 

The presence of diagonal interactions is connected to the isotropy of the interactions in the lattice. This is reflected as a rotationally invariant theory emerging in the continuum limit. Indeed, by restoring the lattice spacing $a$, and though the rescaling of physical variables,
\begin{equation}
S_{\bf r}(t)\rightarrow a^{\frac{(d-z)}{2}}S(t,{\bf r}),~~~{\psi}_{\bf r}(t)\rightarrow a^{\frac{2d-z}{4}}{\psi}(t,{\bf r}),~~~\text{and}~~~\bar{\psi}_{\bf r}(t)\rightarrow a^{\frac{2d-z}{4}}\bar{\psi}(t,{\bf r}),
\label{3.12a}
\end{equation}
together with $\sum_{\bf r}\rightarrow \int \frac{d^dr}{a^d}$, we obtain a continuum theory whose spatial derivatives are given only in terms of rotationally invariant quantities  $S\vec{\nabla}^2 S$, $S(\vec{\nabla}^2)^2 S$, and $\bar{\psi}\vec{\nabla}^2 \psi$,
\begin{equation}
\int dt d^dr \left(\cdots-\frac{1}{a^2}(\tilde{\gamma}\tilde{U}+2d\tilde{U}^2 ) S\vec{\nabla}^2 S-\frac{1}{6}(d+3)\tilde{U}^2 S(\vec{\nabla}^2)^2S-\tilde{U}\bar{\psi}\vec{\nabla}^2 \psi+\cdots\right),
\label{3.13}
\end{equation}
where $\tilde{U}\equiv aU$ and $\tilde{\gamma}\equiv a \gamma$. In the above rescaling we have introduced the dynamical critical exponent $z$, which characterizes the relative scaling between time and spatial correlations. 
Thus, in the Lifshitz point, where the coefficient of the term $S\vec{\nabla}^2 S$ vanishes, we obtain $z=2$.

%%%%%%%%%%%%%%%%%%%%%%%%%%%%%%%%%%%%%%%%%%%%%%%%%%%%%%%%%%%%%%%%%%%%%%%%%%%%%%
\subsection{Mean-Field}

A situation which we will pay special attention is the case of mean-field interactions, where the short-range interactions are replaced by a (weak) interaction involving  
the physical variables of all sites of the lattice. Although the simplicity, this version unveils interesting critical properties and helps to clarify the effect of supersymmetry in the quantum spherical model. The mean-field version is obtained through the replacement    
\begin{eqnarray}
U_{\textbf{r},\textbf{r}'} & \rightarrow & \frac{U}{N},
\label{3.14}
\end{eqnarray}
where $U$ is a constant independent of the site positions. This corresponds to a weak interaction due to the factor $1/N$,  that also guarantees the correct extensivity properties of the free energy. Thus, the first line of the interaction Lagrangian (\ref{3.12}) reduces to
\begin{eqnarray}
{L}_{int}& = & -\frac{1}{N}\left(\gamma U+\frac{U^{2}}{2}\right)\left(\sum_{\textbf{r}}S_{\textbf{r}}\right)^{2}-\frac{U}{N}\left(\sum_{\textbf{r}}\bar{\psi}_{\textbf{r}}\right)\left(\sum_{\textbf{r}'}\psi_{\textbf{r}'}\right).
\label{3.15}
\end{eqnarray}
We see that the bosonic ordering is ferro or anti-ferromagnetic favored as
\begin{eqnarray}
\left(\gamma U+\frac{U^{2}}{2}\right)  >  0~~~\text{or}~~~\left(\gamma U+\frac{U^{2}}{2}\right)  <  0,
\label{3.16}
\end{eqnarray}
respectively. Sec. \ref{S5} is entirely dedicated to this situation.

%%%%%%%%%%%%%%%%%%%%%%%%%%%%%%%%%%%%%%%%%%%%%%%%%%%%%%%

\section{Partition Function}\label{S4}

In this section, we discuss the saddle point computation of the partition function. 
We shall evaluate the partition function in the presence of external fields $H_B$ and $H_F$ through the supersymmetry breaking term,
\begin{equation}
H_B\sum_{\bf r} S_{\bf r}+H_F\sum_{\bf r} \bar{\psi}_{\bf r}\psi_{\bf r}.
\end{equation}
Thus, by taking posteriorly derivatives of the free energy with respect to $H_B$ and $H_F$, we will obtain the usual bosonic order parameter 
$\langle S_{\bf r}\rangle$ and the fermionic condensate $\langle \bar{\psi}_{\bf r}\psi_{\bf r}\rangle$, respectively. 
The finite-temperature partition function is obtained through a Wick rotation to the imaginary time \cite{Kapusta,Das}, though $t=-i\tau$, with $\tau\in [0,\beta]$, 
\begin{eqnarray}
\mathcal{Z} & = &\int\mathcal{D}\Omega\exp\left\{
-\int_0^\beta d\tau\left[ L_E+H_B\sum_{\bf r}S_{\bf r}
+H_F\sum_{\bf r}\bar{\psi}_{\bf r}\psi_{\bf r}\right]\right\},
\label{eq:Z1}
\end{eqnarray}
where the measure $\mathcal{D}\Omega$ corresponds to the integral over all fields as well as over the Lagrange multipliers that implement the supersymmetric constraints,
$\mathcal{D}\Omega\equiv \mathcal{D}S\mathcal{D}F\mathcal{D}\psi\mathcal{D}\bar{\psi}\mathcal{D}\mu\mathcal{D}\gamma\mathcal{D}\xi\mathcal{D}\bar{\xi}$, and $L_E$ is the Euclidean version of Eq. (\ref{2.19}), 
\begin{eqnarray}
L_E&=&\frac{1}{2g}\sum_{\bf r}\dot{S}_{\bf r}^2
-\frac{1}{2}\sum_{\bf r}F_{\bf r}^2+\frac{1}{\sqrt{g}}\sum_{\bf r}\bar{\psi}_{\bf r}\dot{\psi_{\bf r}}
-\sum_{{\bf r},{\bf r}'}U_{{\bf r},{\bf r}'}\left(
S_{\bf r}F_{{\bf r}'} -\bar{\psi}_{\bf r}\psi_{{\bf r}'}\right)\nonumber\\
&-&\gamma\left(\sum_{\bf r}F_{\bf r}S_{\bf r}-\sum_{\bf r}\bar{\psi}_{\bf r}\psi_{\bf r}\right)
+\sum_{\bf r}\bar{\psi}_{\bf r}\xi S_{\bf r}
+\sum_{\bf r}\bar{\xi}\psi_{\bf r} S_{\bf r}
+\mu\left(\sum_{\bf r}S_{\bf r}^2-N\right).
\end{eqnarray}
We remind that, at finite temperature, bosons and fermions have opposite boundary conditions in the imaginary time. Indeed, while the bosonic fields are periodic the fermionic fields are anti-periodic,
\begin{equation}
S_{\bf r}(0)=S_{\bf r}(\beta),~~~F_{\bf r}(0)=F_{\bf r}(\beta),~~~\psi_{\bf r}(0)=-\psi_{\bf r}(\beta),~~~\text{and}~~~\bar{\psi}_{\bf r}(0)=-\bar{\psi}_{\bf r}(\beta),
\label{bc}
\end{equation}
and similarly for the bosonic and fermionic Lagrange multipliers. 

As in the usual classical and quantum spherical models, we can distinguish between two cases, according the way the spherical constraint is implemented. Notice that we are integrating over the Lagrange super multiplier, which means that we are implementing the super spherical constraint strictly (as in (\ref{2.5})). This is in contrast to the so-called mean spherical model, where the constraint is implemented on the average, $\langle \sum_{\bf r}\Phi_{\bf r}^2\rangle=N$. These two cases correspond to distinct ensemble formulations and, for short-range interactions, they are expected to have the same thermodynamic limit. However, it has been pointed out  in \cite{Kastner} that the mean-field version of the classical spherical model exhibits the partial equivalence of ensembles, which is a kind of mild nonequivalence of ensembles. We do not investigate the mean spherical constraint in this work.

The functional integrals over the fields $S,\psi,\bar{\psi}$, and $F$ are all Gaussian and can be straightforwardly performed. These integrations produce
\begin{eqnarray}
\mathcal{Z}=\int \mathcal{D}\mu \mathcal{D}\gamma \mathcal{D}\bar{\xi}
\mathcal{D}\xi\text{e}^{-NS_eff},
\label{34}
\end{eqnarray}
with the effective action given by
\begin{eqnarray}
S_{eff}&\equiv&\frac{1}{2N}\text{Tr}\sum_{\bf q}\ln \left[
-\frac{1}{2g}\frac{\partial^2}{\partial \tau^2}+\mu+
\frac{( \widehat{U}({\bf q})+\gamma)^2}{2}
\right]\nonumber\\
&-&\frac{1}{N}\text{Tr}\sum_{\bf q}\ln\left[
\frac{1}{\sqrt{g}}\frac{\partial}{\partial \tau}+ \widehat{U}({\bf q})+\gamma
+H_F-\frac{1}{2}\xi\mathcal{O}^{-1}_{\bf q}\bar{\xi}
\right]\nonumber\\
&-&\frac{1}{4}\int_0^{\beta}d\tau\frac{H_B^2}{\mu+\frac{\left[\widehat{U}(0)+\gamma\right]^2}{2}}
-\int_0^{\beta}d\tau\mu.
\label{35}
\end{eqnarray}
In this expression $ \widehat{U}({\bf q})$ is the Fourier transform of the interaction $U_{{\bf r},{\bf r}'}\equiv U(|{\bf h}|)$,
\begin{eqnarray}
 \widehat{U}({\bf q})=\sum_{\bf h}U(|{\bf h}|)
\text{e}^{i{\bf q}\cdot{\bf h}},
~~~~\text{with}~~~~
{\bf h}={\bf r}-{\bf r}',
\label{ap8a}
\end{eqnarray}
and the operator $\mathcal{O}_{\bf q}$ is defined as
\begin{eqnarray}
\mathcal{O}_{\bf q}\equiv-\frac{1}{2g}\frac{\partial^2}{\partial\tau^2}+\mu+\frac{1}{2}\left[\widehat{U}({\bf q})+\gamma\right]^2.
\label{ap22a}
\end{eqnarray}
In the effective action (\ref{35}), the trace can be taken, for example, with respect to a ``coordinate" basis labelled by the imaginary time, $|\tau\rangle$, namely, $\text{Tr}\mathcal{O}=\int_0^{\beta}  d\tau \langle\tau| \mathcal{O}|\tau\rangle$, with nondiagonal contributions due to the differential operator $\partial/\partial\tau$. As usual, 
it can be computed by introducing a ``momentum" basis $|n\rangle$, which diagonalizes the operator $\partial/\partial\tau$\footnote{In the case of time independent solutions for $\mu$ and $\gamma$, which we shall consider, it is more convenient to take the trace directly in the momentum basis, $\text{Tr}\mathcal{O}(\partial/\partial\tau)= \sum_n\langle n |\mathcal{O(\partial/\partial\tau)} |n\rangle=\sum_n\mathcal{O}(i\omega_n)$.}, 
\begin{eqnarray}
\frac{\partial}{\partial\tau}|n\rangle=iw_n |n\rangle,~~~\text{with}~~~\langle\tau|n\rangle=\frac{1}{\sqrt{\beta}}\text{e}^{i w_n\tau}.
\label{38}
\end{eqnarray}
The opposite boundary conditions in (\ref{bc}) imply a different spectrum for the frequencies $\omega_n$ for bosons and fermions, i.e., the usual Matsubara frequencies. Explicitly, they are $\omega_n^B=2n\pi/\beta$ for bosons and $\omega_n^F=(2n+1)\pi/\beta$ for fermions,  with $n\in \text{Z}$. This must be taken into account in the evaluation of the trace contributions in the effective action originated from integration over bosonic and fermionic fields. 

The remaining functional integrals in (\ref{34}) can be evaluated through the saddle point method, which becomes exact in the thermodynamic limit $N\rightarrow\infty$. The saddle point equations are determined by the conditions
\begin{eqnarray}
\frac{\delta S_{eff}}{\delta \mu}
=\frac{\delta S_{eff}}{\delta \gamma}
=\frac{\delta S_{eff}}{\delta \xi}
=\frac{\delta S_{eff}}{\delta \bar{\xi}}=0.
\label{36}
\end{eqnarray}
We shall look for time independent saddle point solutions for $\mu,\gamma,\xi$, and $\bar{\xi}$, which can be explicitly evaluated with help of the identity $\delta\text{Tr}\ln \mathcal{A}=\text{Tr} \mathcal{A}^{-1}\delta\mathcal{A}$. Let us start with the two last conditions for the fermionic Lagrange multipliers $\xi$ and $\bar{\xi}$. They are identically satisfied with $\bar{\xi}=\xi=0$. For the first condition, we obtain
{\begin{eqnarray}
0=\frac{\delta S_{eff}}{\delta\mu}=-1+\frac{H_B^2}{4\left[\mu+\frac{(\widehat{U}(0)+\gamma)^2}{2}\right]^2}
+\frac{1}{2N\beta}\sum_{{\bf q}}\sum_{n=-\infty}^{\infty} 2g \frac{1}{(w_n^B)^2+2g\left[
\mu+\frac{(\widehat{U}({\bf q})+\gamma)^2}{2}\right]}.
\label{37}
\end{eqnarray}
The sum over the bosonic Matsubara frequencies can be computed by using
\begin{eqnarray}
\sum_{n=-\infty}^{\infty}\frac{1}{n^2+y^2}=\frac{\pi}{y}
\coth(\pi y),~~~y>0,
\label{40}
\end{eqnarray}
which leads to the constraint
\begin{eqnarray}
1=\frac{H_B^2}{4\left[\mu+\frac{1}{2}(\widehat{U}(0)+\gamma)^2\right]^2}
+\frac{1}{2N}\sum_{\bf q}\frac{g}{w_{\bf q}^B}\coth\left(
\frac{\beta}{2}w_{\bf q}^B
\right),
\label{41}
\end{eqnarray}
with the bosonic frequency defined as
\begin{eqnarray}
\left(w_{\bf q}^B\right)^2\equiv 2g\left\{\mu+\frac{1}{2}\left[\widehat{U}({\bf q})+\gamma\right]^2\right\}.
\label{42}
\end{eqnarray}
The second saddle point condition yields to
\begin{eqnarray}
0=\frac{\delta S_{eff}}{\delta \gamma}
&=&\frac{H_B^2}{4\left[\mu+\frac{(\widehat{U}(0)+\gamma)^2}{2}\right]^2}~
\left[\widehat{U}(0)+\gamma\right]
+\frac{1}{2N}\sum_{\bf q}\frac{g}{w_{\bf q}^B}~
[\widehat{U}({\bf q})+\gamma]\coth\left(
\frac{\beta}{2}w_{\bf q}^B
\right)\nonumber\\
&-&\frac{1}{N\beta}\sum_{{\bf q}}\sum_{n=-\infty}^{\infty}\frac{1}{\frac{1}{\sqrt{g}}iw_n^F+
\widehat{U}({\bf q})+\gamma+H_F}.
\label{43}
\end{eqnarray}
To evaluate the sum in the last line we note that it can be written as 
\begin{eqnarray}
\sum_{n=-\infty}^{\infty}\frac{1}{\frac{1}{\sqrt{g}}iw_n^F+
\widehat{U}({\bf q})+\gamma+H_F}=\left(\widehat{U}({\bf q})+\gamma+H_F \right)
 \sum_{n=-\infty}^{\infty}\frac{1}{\frac{1}{g}(w_n^F)^2+
\left(\widehat{U}({\bf q})+\gamma+H_F\right)^2}.
\label{44}
\end{eqnarray}
The direct use of 
\begin{equation}
\sum_{n=-\infty}^{\infty}\frac{1}{(2n+1)^2+y^2}=\frac{\pi}{2y}
\tanh\left(\frac{\pi y}{2}\right),
\end{equation}
provides the second constraint,
\begin{eqnarray}
0&=&\frac{H_B^2}{4\left[\mu+\frac{(\widehat{U}(0)+\gamma)^2}{2}\right]^2}~
\left[\widehat{U}(0)+\gamma\right]
+\frac{1}{2N}\sum_{\bf q}\frac{g}{w_{\bf q}^B}~
[\widehat{U}({\bf q})+\gamma]\coth\left(
\frac{\beta}{2}w_{\bf q}^B
\right)\nonumber\\
&-&\frac{1}{N}\sum_{\bf q}\frac{g}{2w_{\bf q}^F}[\widehat{U}({\bf q})+\gamma+H_F]
\tanh\left(\frac{\beta}{2}w_{\bf q}^F\right),
\label{46}
\end{eqnarray}
where the fermionic frequency, incorporating the external field, is defined as
\begin{eqnarray}
(w_{\bf q}^F)^2= g\left[\widehat{U}({\bf q})+\gamma+H_F\right]^2.
\label{45}
\end{eqnarray}

Equations (\ref{41}) and (\ref{46}) together with the effective action (\ref{35}) (which is essentially the free energy) are the basic relations to study the critical behavior of the model. As we shall discuss, they determine the values of $\mu$ and $\gamma$ compatible with the existence of a critical point. Although the partition function can be exactly computed in the thermodynamic limit for an arbitrary interaction $\widehat{U}({\bf q})$, the relations between the parameters in (\ref{41}) and (\ref{46}) are given in terms of multidimensional integrals $\int d^d{\bf q}$, which makes the analysis a little involved, often requiring numerical calculations. 
A complete analysis of the critical behavior for the case of short-range interactions will be reported elsewhere. In what follows, we shall pursue a detailed study of the mean-field critical behavior, where all the relations are given explicitly. This analysis is interesting in its own and provides insights about phase transitions in the model, unveiling nice critical properties in both cases of zero and finite temperature.

%%%%%%%%%%%%%%%%%%%%%%%%%%%%%%%%%%%%%%%%%%%%%%%%%%%5
 
\subsection{Saddle Point Solutions and Supersymmetry Breaking} 

In the previous discussion we had some elements which break the supersymmetry, namely, the external fields, $H_B$ and $H_F$, and the temperature $k_BT=1/\beta$. 
Supersymmetry is broken at the finite temperatures once we have distinct thermal distributions for bosons and fermions. However, at zero temperature and in the absence of the external fields, any supersymmetry breaking must be spontaneously. In terms of the saddle point parameters, this happens whenever it is possible to find a solution of (\ref{41}) and (\ref{46}) with $\mu\neq 0$.    
As the first hint of this, we note that the bosonic ($\omega_{\bf q}^B$) and fermionic ($\omega_{\bf q}^F$) frequencies are the same only when $\mu=0$, independent of the value of $\gamma$. Concretely, we shall look at the ground state energy that is a function of $\mu$ and $\gamma$, as the supersymmetry requires the vanishing of the ground state energy. 

We can compute the ground state energy from the effective action (\ref{35}), which is essentially the free energy of the model.  
By proceeding similarly to the previous calculation, we find the following result for the free energy with $H_B=H_F=0$,
\begin{equation}
f=\frac{1}{\beta}S_{eff}=-\mu+\frac{1}{\beta N}\sum_{\bf q}  \ln\left[2\sinh\left(\frac{\beta\omega_{\bf q}^B}{2}\right)\right]
-\frac{1}{\beta N}\sum_{\bf q}  \ln\left[2\cosh\left(\frac{\beta\omega_{\bf q}^F}{2}\right)\right].
\label{8.1}
\end{equation} 
By taking the zero-temperature limit, the free energy reduces to the ground state energy $E_0$,
\begin{eqnarray}
\frac{E_0}{N}&=&-\mu+\frac{1}{N}\sum_{\bf q} (\omega_{\bf q}^B-\omega_{\bf q}^F)\nonumber\\
&=&-\mu+\frac{1}{N}\sum_{\bf q} 
\left\{
\left[2g\left(\mu+\frac{1}{2}\left(\widehat{U}({\bf q})+\gamma\right)^2\right)\right]^{\frac{1}{2}}
-\left[2g\left(\frac{1}{2}\left(\widehat{U}({\bf q})+\gamma\right)^2\right)\right]^{\frac{1}{2}}
\right\},
\label{8.2}
\end{eqnarray}
which vanishes only when $\mu=0$, independent of $\gamma$. In the next section, we explore these points in the mean-field version of the supersymmetric model.

%%%%%%%%%%%%%%%%%%%%%%%%%%%%%%%%%%%%%%%%%%%%%%%%%%%%%%%%%%%%%%%%%%%%%%%%%%%%
\section{Mean-Field Critical Behavior}\label{S5}

As discussed previously, the mean-field version of the model is obtained from the replacement in (\ref{3.14}). In terms of the Fourier transform of the interaction, this corresponds to
\begin{equation}
\widehat{U}({\bf q})\rightarrow U\delta_{{\bf q},0}.
\label{FourierMF}
\end{equation}
Thus the frequencies $w_{{\bf q }}^B$ and $w_{{\bf q }}^F$ split in two parts, the part containing the zero mode, ${\bf q}=0$, and the part containing the remaining ones, 
\begin{eqnarray}
\left(w_{{\bf q }= 0}^B\right)^2=2g\left[\mu+\frac{\left(U+\gamma\right)^2}{2}\right]
~~~\text{and}~~~\left(w_{\bf q \neq 0}^B\right)^2
=2g\left[\mu+\frac{\gamma^2}{2}\right]
\label{48}
\end{eqnarray}
and
\begin{eqnarray}
(w_{{\bf q}=0}^F)^2= g\left(U+\gamma+H_F
\right)^2 ~~~~\text{and}~~~
(w_{{\bf q}\neq 0}^F)^2= g\left(\gamma+H_F
\right)^2.
\label{50}
\end{eqnarray}
The constraint equations (\ref{41}) and (\ref{46}) reduce to 
\begin{eqnarray}
1=\frac{H_B^2}{4\left[\mu+\frac{1}{2}(U+\gamma)^2\right]^2}
+\frac{1}{2N}\frac{g}{w_{{\bf q }= 0}^B}\coth\left(
\frac{\beta}{2}w_{{\bf q }= 0}^B
\right)
+\frac{(N-1)}{2N}\frac{g}{w_{{\bf q}\neq 0}^B}\coth\left(
\frac{\beta}{2}w_{{\bf q}\neq 0}^B
\right)
\label{47}
\end{eqnarray}
and
\begin{eqnarray}
0&=&\frac{H_B^2}{4\left[\mu+\frac{(U+\gamma)^2}{2}\right]^2}~
\left[U+\gamma\right]
+\frac{1}{2N}\frac{g}{w_{{\bf q}=0}^B}~
[U+\gamma]\coth\left(
\frac{\beta}{2}w_{{\bf q}=0}^B
\right)\nonumber\\
&+&\frac{(N-1)}{2N}\frac{g}{w_{{\bf q}\neq0}^B}~
\gamma\coth\left(
\frac{\beta}{2}w_{{\bf q}\neq0}^B
\right)
-\frac{1}{2N}\frac{g}{w_{{\bf q}=0}^F}(U+\gamma+H_F)
\tanh\left(\frac{\beta}{2}w_{{\bf q}=0}^F\right)\nonumber\\
&-&\frac{(N-1)}{2N}\frac{g}{w_{{\bf q}\neq0}^F}(\gamma+H_F)
\tanh\left(\frac{\beta}{2}w_{{\bf q}\neq0}^F\right).
\label{49}
\end{eqnarray}
A phase transition can be detected in these expressions by identifying certain values of the involved parameters corresponding to a point of nonanalyticity emerging in the thermodynamic limit\footnote{This mechanism is similar to what happens in a Bose-Einstein condensation \cite{Gunton}.}. Accordingly, an order parameter is expected to exhibit different behavior as we cross such a critical point.  As the phase transition can be governed by thermal or quantum fluctuations, we should analyze the corresponding critical behaviors separately, starting with the zero-temperature case where the phase transition is driven by quantum fluctuations.

%%%%%%%%%%%%%%%%%%%%%%%%%%%%%%%%%%%%%%%%%%%%%%%%%%%%%%%%%%%%%
\subsection{Quantum Critical Behavior}

We have to analyze the behavior of the constraints (\ref{47}) and (\ref{49}) in the zero-temperature limit, 
which enable us to obtain the parameters $\mu$ and $\gamma$ as a function of $g$, $H_B$, and $H_F$. The expression (\ref{47}) for $\beta \rightarrow \infty$ is reduced to
\begin{eqnarray}
1=\frac{H_B^2}{4\left[\mu+\frac{1}{2}(U+\gamma)^2\right]^2}
+\frac{1}{2N}\frac{g}{\sqrt{2g\left[\mu+\frac{\left(U+\gamma\right)^2}{2}\right]}}+
\frac{(N-1)}{2N}\frac{g}{{\sqrt{2g\left[\mu+\frac{\gamma^2}{2}\right]}}}.
\label{cb1}
\end{eqnarray}
For the equation (\ref{49}), turning off the external field and taking the thermodynamic limit we get
\begin{equation}
\frac{1}{\sqrt{\mu+\frac{\gamma^2}{2}}}=\frac{1}{\sqrt{\frac{\gamma^2}{2}}},
\end{equation}
which implies that $\mu=0$ independent of the value of $\gamma$. This result shows that  supersymmetry is not spontaneously broken since $\mu\neq 0$ does not correspond to a saddle point solution. 

By setting $\mu=0$ in (\ref{cb1}) and solving it for $g$, with $H_B= H_F=0$, it follows
\begin{eqnarray}
\frac{1}{\sqrt{g}}=\frac{1}{2|\gamma|} + \frac{|\gamma| -|U+\gamma|}{2 N |\gamma||U+\gamma|},
\label{cb03}
\end{eqnarray}
valid for $|\gamma| > |U+\gamma|$. We see that this condition is achieved only when $U$ and $\gamma$ have opposite signs.
The Eq. (\ref{cb03}) allows us to write an expression for $\gamma$ as a function of $g$ and $N$, which exhibits a  point of nonanalyticity in the  
thermodynamic limit,   
\begin{eqnarray}
\gamma=\left\{
\begin{array}
[c]{ccc}
\pm |U| & \text{for}  & \sqrt{g}<2|U|\\
\pm \frac{\sqrt{g}}{2} &  \text{for} & \sqrt{g}>2|U|
\end{array}
\right. ,
\label{cb04}
\end{eqnarray}
with the $(+)$ sign corresponding to the case $U<0$ and the $(-)$ sign corresponding to $U>0$. The general pattern as $N$ is increased is shown in Fig. \ref{fig1}.
Therefore, this analysis shows that there is a zero-temperature critical point at $\sqrt{g}=\sqrt{g_c}\equiv 2|U|$, such that the model exhibits a quantum phase transition without breaking supersymmetry. The corresponding parameter space defined by the saddle point solution is illustrated in Fig. \ref{zeroT}.
\begin{figure}[h]
\includegraphics[width=7cm,height=6cm]{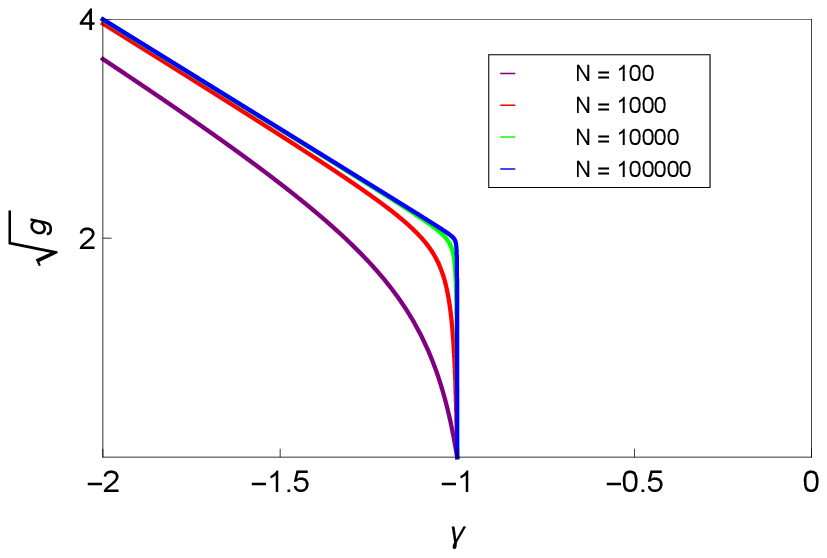}
\includegraphics[width=7cm,height=6cm]{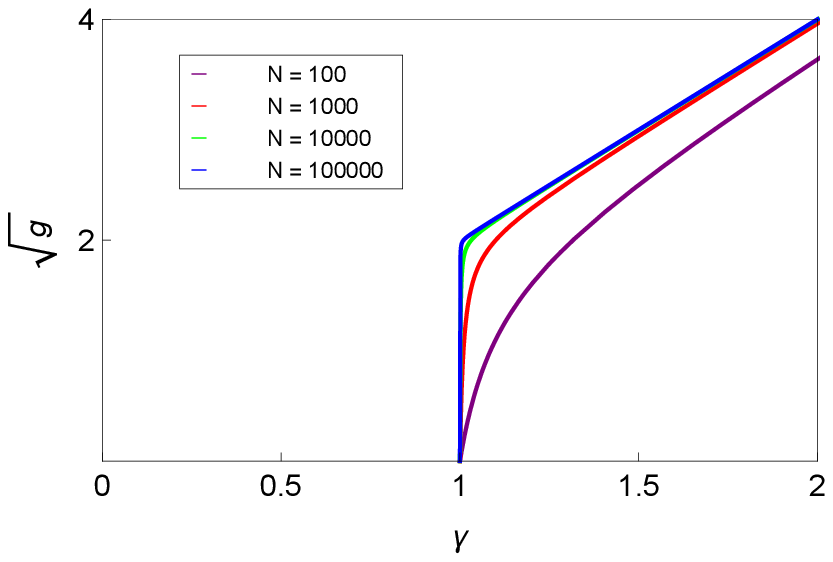}
\caption{Formation of a point of nonanalyticity dictated by Eq. (\ref{cb03}) as $N$ is increased. The plots are with $U\equiv1$ in the left panel and $U\equiv-1$ in the right panel.}
\label{fig1}
\end{figure}
\begin{figure}[h]
\includegraphics[width=7cm,height=6cm]{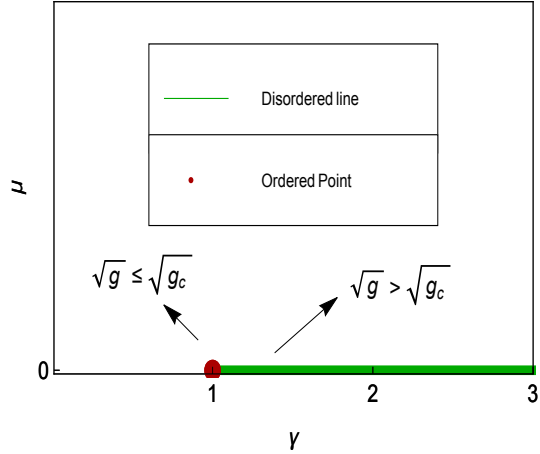}
\caption{Zero-temperature parameter space for the case $U<0$, ($U\equiv-1$). There is a quantum phase transition without breaking supersymmetry.}
\label{zeroT}
\end{figure}

%%%%%%%%%%%%%%%%%%%%%%%%%%%%%%%%%%%%%%%%%%%%%%%%%%%5
\subsubsection{Magnetization, Fermionic Condensate, and Susceptibility} 

Thermodynamic quantities can be obtained from the free energy, (\ref{8.1}), which, in the presence of the external fields $H_B$ and $H_F$, reads,
\begin{eqnarray}
f&=&-\frac{H_B^2}{4\left[\mu+\frac{1}{2}\left(U(0)+\gamma\right)^2\right]}-\mu+\frac{1}{\beta N}\sum_{\bf q}\ln \left[
2\sinh\left(\frac{\beta w_{\bf q}^B}{2}\right)\right]\nonumber\\
&-&\frac{1}{\beta N}\sum_{\bf q}\ln \left[
2\cosh\left(\frac{\beta w_{\bf q}^F}{2}\right)\right].
\label{cb09}
\end{eqnarray}
In the mean-field version it takes the form
\begin{eqnarray}
f&=& -\frac{H_B^2}{4\left[\frac{1}{2}\left(U+\gamma\right)^2\right]}
+\frac{1}{\beta N}\left\{  \ln \left[
2\sinh\left(\frac{\beta w_0^B}{2}\right)\right]+ (N-1)\ln \left[
2\sinh\left(\frac{\beta w_{{\bf q}\neq 0}^B}{2}\right)\right] \right\}\nonumber\\
&-&\frac{1}{\beta N}\left\{\ln \left[
2\cosh\left(\frac{\beta w_0^F}{2}\right)\right]+ (N-1)\ln \left[
2\cosh\left(\frac{\beta w_{{\bf q}\neq 0}^F}{2}\right)\right]\right\}.
\label{cb10}
\end{eqnarray}
As introduced in the beginning of Sec. \ref{S4}, we shall investigate the usual bosonic magnetization, 
\begin{eqnarray}
m_B&\equiv&\left<\frac{1}{N}\sum_{\bf r}S_{\bf r}\right>=
\frac{1}{N\beta}\frac{\partial \ln \mathcal{Z}}{\partial H_B}=-\frac{\partial f}{\partial H_B}
\label{cb12a}
\end{eqnarray}
and the fermionic condensate,
\begin{eqnarray}
\mathcal{C}_F&\equiv&\left<
\frac{1}{N}\sum_{\bf r}\bar{\psi}_{\bf r}\psi_{\bf r}\right>=
\frac{1}{N\beta}\frac{\partial \ln \mathcal{Z}}{\partial H_F}=-\frac{\partial f}{\partial H_F}.
\label{cb12b}
\end{eqnarray}

In the zero-temperature limit, the bosonic magnetization is given by
\begin{equation}
m_B=\frac{H_B}{\left(\gamma+U\right)^2}.
\label{cb12}
\end{equation}
For $\sqrt{g}>\sqrt{g_c}$, the quantity $(\gamma+U)$ is always different from zero and therefore when 
$H_B=0$, the magnetization is zero. On the other hand, when $\sqrt{g}<\sqrt{g_c}$, we have $(\gamma+U)=0$ and the magnetization gives an indeterminacy when $H_B=0$. 
To handle this, we use the constraint, 
\begin{eqnarray}
1=\frac{H_B^2}{4\left[\frac{1}{2}(U+\gamma)^2\right]^2}+\frac{1}{2N}\frac{g}{w_0^B}
+\frac{(N-1)}{2N}\frac{g}{w_{{\bf q}\neq 0}^B},
\label{cb13}
\end{eqnarray}
in conjunction with (\ref{cb12}) to write the magnetization without explicit dependence on the external field $H_B$. After this, and considering the thermodynamic limit, we get 
\begin{eqnarray}
1&=&m_B^2+ \frac{\sqrt{g}}{2|\gamma|}.
\label{cb14}
\end{eqnarray}
As we are below the critical point, where $|\gamma|=|U|= \sqrt{g_c}/2$, this relation yields
\begin{eqnarray}
m_B&=\pm&\left(\frac{\sqrt{g_c}-\sqrt{g}}{\sqrt{g_c}}\right)^{\frac{1}{2}},
\label{cb15}
\end{eqnarray}
which shows that the quantum critical exponent $\beta_g$ of the bosonic order parameter is $\beta_g=1/2$.

Now let us discuss the behavior of the fermionic condensate. By computing the derivative with respect to $H_F$ as indicated in Eq. (\ref{cb12b}), we obtain 
\begin{eqnarray}
\mathcal{C}_F=\frac{\sqrt{g}}{2}\text{sign}(\gamma+H_F).
\end{eqnarray}
By turning off $H_F$, we see that the fermionic condensate is nonvanishing no matter in what phase we are, behaving uniformly both above and below the critical point. 

From the Eq. (\ref{cb12}) we obtain the bosonic susceptibility,
\begin{eqnarray}
\chi_B=\frac{\partial m_B}{\partial H_B} = \frac{1}{(\gamma+U)^2},
\label{cb16}
\end{eqnarray}
which diverges for $\sqrt{g}<\sqrt{g_c}$ because of $(\gamma+U)=0$ \footnote{The same behavior is also observed in the case of the classical spherical model below the critical temperature.}. For $\sqrt{g}>\sqrt{g_c}$
we have  $\gamma=\pm\sqrt{g}/2$ and $U= \pm\sqrt{g_c}/2$ and, taking into account that $\gamma$ and $U$ should have 
opposite signs, we obtain
\begin{eqnarray}
\chi_B\propto\left(\sqrt{g}-\sqrt{g_c}\right)^{-2},
\label{cb17}
\end{eqnarray}
giving a new bosonic quantum critical exponent $\gamma_g=2$. 

As mentioned in the introduction, the classical spherical model has a pathological behavior at low temperature, with the entropy diverging as $T\rightarrow 0$ $(S\sim \ln T)$. 
It is interesting to investigate the low-temperature behavior of the entropy in our model. In the thermodynamic limit, it is
\begin{eqnarray}
\frac{1}{k_B}s&=&\beta^2\frac{\partial f}{\partial \beta}\nonumber\\
&=&-\ln\left[2\sinh\left(\frac{\beta w_{{\bf q}\neq 0}^B}{2}\right)\right]
+\ln\left[2\cosh\left(\frac{\beta w_{{\bf q}\neq 0}^F}{2}\right)\right]\nonumber\\
&+&\beta\left[
\frac{w_{{\bf q}\neq 0}^B}{2}
\coth\left(\frac{\beta w_{{\bf q}\neq 0}^B}{2}\right)
-\frac{w_{{\bf q}\neq 0}^F}{2}
\tanh\left(\frac{\beta w_{{\bf q}\neq 0}^F}{2}\right)
\right].
\label{cb18}
\end{eqnarray}
For $H_B=H_F=0$, the bosonic and fermionic frequencies are the same, $w_{{\bf q}\neq 0}^B=w_{{\bf q}\neq 0}^F$, and it is easy to verify that the entropy vanishes as $T\rightarrow 0$.

%%%%%%%%%%%%%%%%%%%%%%%%%%%%%%%%%%%%%%%%%%%%%
\subsection{Critical Behavior at Finite Temperature}

The whole analysis here is similar to that one of the previous section. However, in this situation the supersymmetry is broken by the temperature and we can expect to find saddle point solutions with $\mu\neq 0$. 
The critical behavior governed by thermal fluctuations is obtained by considering the thermal energy higher than all the quantum energy scales (frequencies). In this situation, we can expand the coth's for small argument in Eq. (\ref{47}) and retain only the leading term, $\coth(x)= \frac{1}{x}+O(x)$, which effectively contributes at the critical point, 
\begin{eqnarray}
1=\frac{H_B^2}{4\left[\mu+\frac{1}{2}(U+\gamma)^2\right]^2}
+\frac{1}{2N \beta}\frac{1}{\left[\mu+\frac{\left(U+\gamma\right)^2}{2}\right]}
+
\frac{(N-1)}{2N\beta}\frac{1}{\left[\mu+\frac{\gamma^2}{2}\right]}.
\label{cb19}
\end{eqnarray}
This expression exhibits a critical behavior only for $\mu\leq 0$, with the critical point $\mu=-(U+\gamma)^2/2$.
Putting $H_B=H_F=0$ and taking the thermodynamic limit, the second constraint given in Eq. (\ref{49}) leads to
\begin{equation}
g=\frac{2}{\beta^2\left(\mu+\frac{\gamma^2}{2}\right)},
\label{g}
\end{equation}
where we have also expanded the hyperbolic functions. Notice that $\mu+\frac{\gamma^2}{2}>0$ is the requirement for the bosonic frequencies to be real and then is always fulfilled. Thus, for any $\gamma,\mu$ and $\beta$ satisfying Eq. (\ref{cb19}), there is a value of  $g$ given above that satisfies the saddle point condition (\ref{49}).

Now we can investigate the arising of a point of nonanalyticity in (\ref{cb19}) as $N$ is increased, by solving it for $\gamma$ as a function of $\beta$ keeping both $\mu$ and $N$ fixed. The general pattern is shown in Fig. \ref{fig3}. Fig. \ref{fig4} shows its behavior in the thermodynamic limit for distinct values of $\mu$.
In general, the Eq. (\ref{cb19}) in the thermodynamics limit and with $H_B=H_F=0$ yields, 
\begin{eqnarray}
\gamma=\left\{
\begin{array}
[c]{ccc}
\pm (|U| +\sqrt{2|{\mu}}|)&~~ \text{for}~~  & k_BT<k_BT_c\\
\pm \left(k_BT+2|\mu|\right)^{\frac{1}{2}}&~~  \text{for}~~ & k_BT>k_BT_c
\end{array}
\right. ,
\label{cb21}
\end{eqnarray}
with the $(+)$ sign corresponding to $U < 0$ and the $(-)$ sign corresponding to $U > 0$. The two solutions define a critical point, $k_BT= k_BT_c\equiv U^2+2|U|\sqrt{2|\mu|}$.
The corresponding parameter space is shown in Fig. \ref{Tfinite}.

\begin{figure}[h]
\includegraphics[width=7cm,height=6cm]{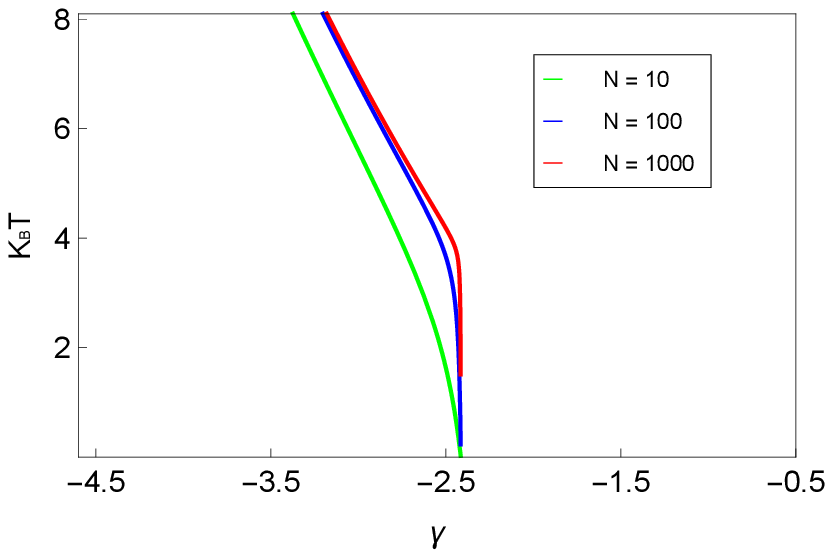}
\includegraphics[width=7cm,height=6cm]{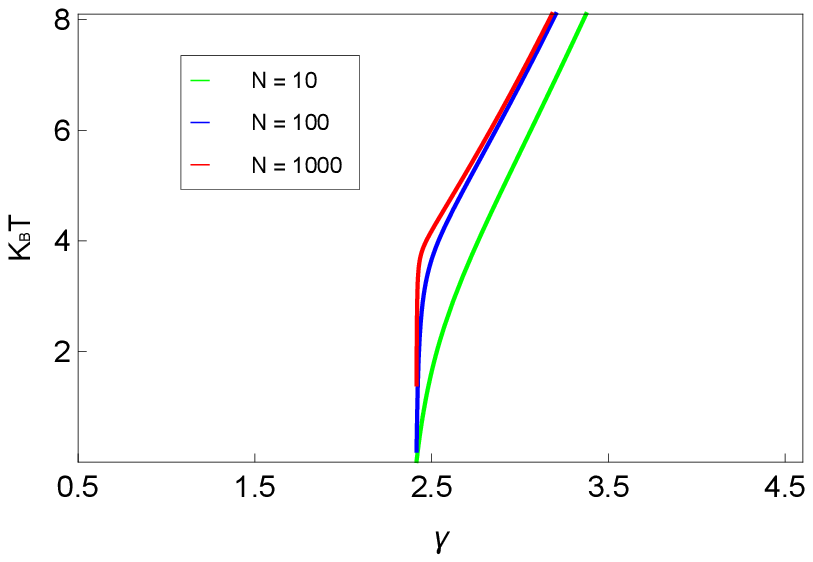}
\caption{Formation of a point of nonanalyticity in the case of finite temperature as $N$ is increased. The plots are with $\mu=-1$ and $H_B=0$, and with
$U\equiv1$ in the left panel and $U\equiv-1$ in the right panel. In this case $k_BT_c= 3.82843$, in accordance with Eq. (\ref{cb21})}
\label{fig3}
\end{figure}

\begin{figure}[h]
\includegraphics[width=7cm,height=6cm]{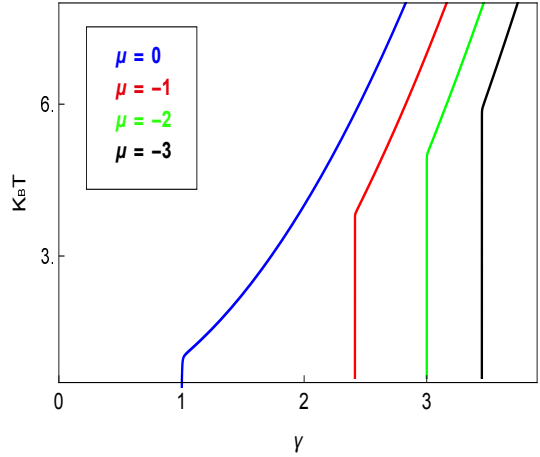}
\caption{The point of nonanaliticity arising in Eq. (\ref{cb19}) in the thermodynamic limit for different values of $\mu$.}
\label{fig4}
\end{figure}

\begin{figure}[h]
\includegraphics[width=7cm,height=6cm]{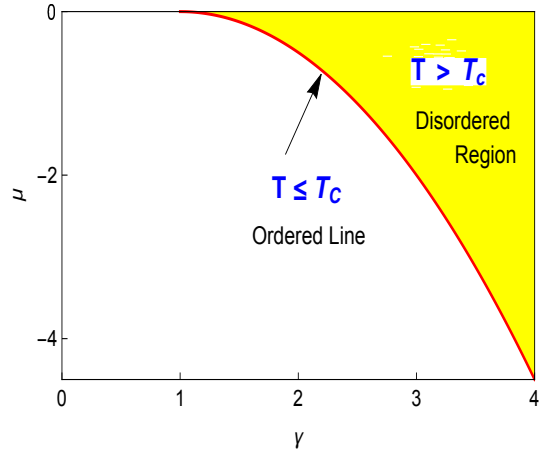}
\caption{The finite-temperature parameter space, with $U\equiv -1$.}
\label{Tfinite}
\end{figure}

It is interesting to collect the results for the critical behavior in the cases of zero and finite temperature by constructing a phase diagram $k_BT\times g$, with $\mu=0$. It is shown in Fig. \ref{fig2}. The critical line is given by 
\begin{eqnarray}
1=\frac{1}{2}\frac{\sqrt{g}}{|U|}\coth\left[\frac{\beta}{2}\sqrt{g}|U|\right].
\label{cb08}
\end{eqnarray}

\begin{figure}[h]
\includegraphics[width=7cm,height=6cm]{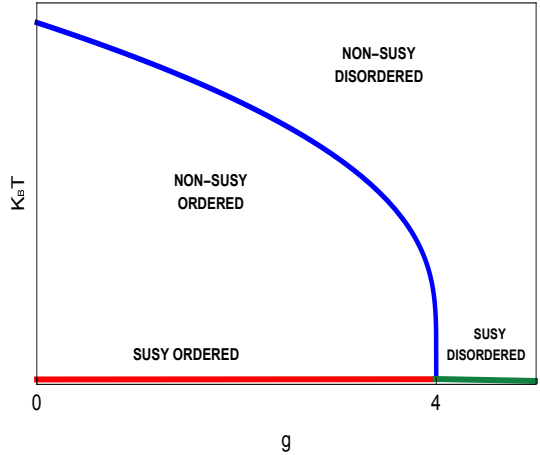}
\caption{Phase diagram of the mean-field version of the supersymmetric quantum spherical model, with $\mu=0$ and $\gamma_c=+|U|\equiv1$. 
The quantum critical point occurs at  $g_c=4|U|^2=4$.}
\label{fig2}
\end{figure}

%%%%%%%%%%%%%%%%%%%%%%%%%%%%%%%%%%%%%%%%%%%%
\subsubsection{Magnetization, Fermionic Condensate, and Susceptibility} 

The magnetization, Eq. (\ref{cb12a}), for the case of finite temperature is
\begin{eqnarray}
m_B= \frac{H_B}{2\left[\mu+\frac{1}{2}(U+\gamma)^2\right]}.
\label{cb22}
\end{eqnarray}
For $k_BT>k_BT_c$, the factor $\mu+\frac{1}{2}(U+\gamma)^2\neq 0$ and then $m_B=0$ for $H_B=0$. 
For $k_BT<k_BT_c$, there is an indeterminacy in $H_B=0$ that can be settled up by means of the constraint (\ref{cb19}). By eliminating the explicit 
dependence on the external field $H_B$ we obtain that, in the thermodynamic limit, the bosonic magnetization satisfies,
\begin{eqnarray}
1=m_B^2+\frac{1}{2\beta}\frac{1}{\left(\mu+\frac{\gamma^2}{2}\right)},
\label{cb24}
\end{eqnarray}
which leads to the following solution 
\begin{eqnarray}
m_B= \pm \left(\frac{T_c-T}{T_c}\right)^{\frac{1}{2}},
\label{cb25}
\end{eqnarray}
where we have used (\ref{cb21}) to write $\gamma$ in terms of the critical temperature.  As in the case of zero temperature, we obtain the usual mean-field critical exponent $\beta_T=1/2$.

The  Eq. (\ref{cb12b}) provides the behavior of fermionic condensate,
\begin{eqnarray}
\mathcal{C}_F=\frac{1}{4} g \beta(\gamma+H_F)+O(\beta^3).
\label{cb23}
\end{eqnarray}
We remember that $\gamma$ has different behaviors above and below the critical point, according to the Eq. (\ref{cb21}), and
$g$ is given in Eq. (\ref{g}). For $k_BT<k_BT_c$, the parameter 
$\gamma$ does not depend on the temperature, such that for $H_F=0$ we obtain 
\begin{eqnarray}
\mathcal{C}_F=\text{sign}({\gamma})\frac{T}{T_c}\left(|U|
+\sqrt{2|\mu|}\right).
\end{eqnarray}
In the case of $k_BT>k_BT_c$, the fermionic condensate is 
\begin{eqnarray}
\mathcal{C}_F=\text{sign}({\gamma})\sqrt{k_BT+2|\mu|}.
\end{eqnarray}
Thus, contrarily to the zero-temperature case, at  finite temperatures the fermionic condensate is sensitive to the phase transition in the sense that its temperature dependence changes as we cross the critical point.

The derivative of the Eq. (\ref{cb22}) with respect to the external field $H_B$ gives the bosonic susceptibility,
\begin{eqnarray}
\chi_B=\frac{\partial m_B}{\partial H_B}=\frac{1}{2\left[\mu +\frac{1}{2}(U+\gamma)^2\right]}.
\end{eqnarray} 
For $k_BT< k_BT_c$, it always diverges since $\mu =-\frac{1}{2}(U+\gamma)^2$.
As mentioned before, for $k_BT> k_BT_c$, the parameter $\gamma$ is a function of $\mu$ and $T$, given in Eq. (\ref{cb21}), and writing $|U|$ in terms of $\mu$ and $T_c$ as
\begin{eqnarray}
|U| = -\sqrt{2|\mu|}+\sqrt{2|\mu|+\frac{1}{\beta_c}},
\end{eqnarray}
we obtain $\chi_B$ in terms of the parameter $\mu$,
\begin{eqnarray}
\chi_B=\left\{\left[\left(2|\mu|+\frac{1}{\beta}\right)^{\frac{1}{2}}-\left(2|\mu|+\frac{1}{\beta_c}\right)^{\frac{1}{2}}\right]^2+2\sqrt{2|\mu|}
\left[\left(2|\mu|+\frac{1}{\beta}\right)^{\frac{1}{2}}-\left(2|\mu|+\frac{1}{\beta_c}\right)^{\frac{1}{2}}\right]\right\}^{-1}.
\end{eqnarray}
By expanding around the critical point this becomes
\begin{eqnarray}
\chi_B&=&\left[\sqrt{2|\mu|}\left(\frac{a}{k_BT}\right)\left(\frac{T-T_c}{T_c}\right)+\left(\frac{a}{k_BT}\right)^2\left(\frac{T-T_c}{T_c}\right)^2
+O\left(\frac{T-T_c}{T_c}\right)^3\right]^{-1},
\label{sc}
\end{eqnarray}
with the parameter $a$ defined as
\begin{eqnarray}
a\equiv\frac{\left(\frac{1+2\beta_c|\mu|}{\beta_c}\right)^{\frac{1}{2}}}{2\beta_c\left(1+2\beta_c|\mu|\right)}.
\end{eqnarray}
The expression (\ref{sc}) shows an interesting feature. Note that for $\mu\neq 0$ ($\mu<0$),
\begin{eqnarray}
\chi_B\propto\left(\frac{T-T_c}{T_c}\right)^{-1},
\end{eqnarray}
we recover the mean-field critical exponent, $\gamma_T=1$. On the other hand, for $\mu=0$, 
the dominant term provides
\begin{eqnarray}
\chi_B\propto\left(\frac{T-T_c}{T_c}\right)^{-2},
\end{eqnarray}
defining a new critical exponent, $\gamma_T(\mu=0)=2$, as in the case of zero temperature.

%%%%%%%%%%%%%%%%%%%%%%%%%%%%%%%%%%%%%%%%%%%%%%%%%%%%%%%
\section{Final Remarks}\label{S6}

We conclude this work by summarizing the main points of the paper. The superspace construction of the model is appropriate to safely handle the constraint structure in compliance with the supersymmetry requirements. On the other hand, the on-shell formulation obtained after integrating out the auxiliary degree of freedom enables a clearer visualization of the type of interactions present in the model. In this context, we briefly discussed the structure arising by assuming that $U_{{\bf r}, {\bf r}'}$ is restricted to first-neighbors. The possibility for competing interactions in the bosonic sector, adds to the model a potential to exhibit a rich critical behavior with modulated phases, in addition to the ordered and disorder ones. In this case, a Lifshitz point is expected at the meeting point of such phases. This analysis provides good perspectives for further studies on the model and it is currently under investigation. 

After determining the saddle point equations we set out to study the critical behavior of the model in the case of mean-field interactions, where $U_{{\bf r}, {\bf r}'} \rightarrow U/N$. In addition to simplifying the saddle point equation, it provides an interesting critical behavior. At zero temperature the saddle point equations requires $\mu=0$,  ensuring that the supersymmetry is not spontaneously broken. With this condition we find a critical behavior whenever $\gamma$ and $U$ have opposite signs, providing a phase transition without breaking supersymmetry. The usual magnetization exhibits a typical mean-field critical exponent, but the susceptibility is characterized by a new critical exponent $\gamma_{g}=2$. The fermionic condensate behaves uniformly as we cross the critical point, being therefore insensitive to the quantum phase transition. 

At finite temperature, the thermal fluctuations are responsible for breaking the supersymmetry. In this situation, the model exhibits a critical behavior for $\mu\leq0$ and when  $\gamma$ and $U$ have opposite signs, as in the case of zero temperature. 
Under these conditions we find that the model undergoes a phase transition at a critical temperature $T_c$.  
It is interesting that the model reveals different critical behaviors according to the values of $\mu$, in the sense that the susceptibility is governed by distinct critical exponents. For $\mu=0$, we obtain the same exponent as the zero-temperature case, $\gamma_T=2$. For $\mu<0$, we recover the usual mean-field value $\gamma_T=1$. 
Regardless the value of $\mu$, the magnetization is characterized by the mean-field critical exponent $\beta=1/2$, as in the case of zero temperature. However, contrarily to the case of zero temperature, the fermionic condensate is sensitive to the thermal phase transition, exhibiting different temperature dependence above and below the critical temperature. These results suggest that the mean-field interactions are too weak to break the condensate in both cases of zero and finite temperature. In a study in progress,
we expect to obtain more precise conclusions about the behavior of the fermionic condensate in the case of short-range interactions.

%%%%%%%%%%%%%%%%%%%%%%%%%%%%%%%%%%%%%%%%%%%%%%%%%%%%%

\section{Acknowledgments}

We would like to thank Carlos Hernaski and Marcelo Gomes for very helpful discussions and comments on the manuscript. We acknowledge the financial support of Brazilian agencies CAPES, CNPq and Funda\c c\~ao Arauc\'aria.

%%%%%%%%%%%%%%%%%%%%%%%%%%%%%%%%%%%%%%%%%%%%%%%%%%%%%%%%%%
%%%%%%%%%%%%%%%%%%%%%%%%%%%%%%%%%%%%%%%%%%%%%%%%%%%%%%%%%

\appendix

\section{Superspace Formalism}

In this appendix, we recall some relevant properties of the superspace formalism in supersymmetric quantum mechanics \cite{Shifman,Bagchi}, which are useful for the construction of the model discussed in this work. We also clarify the meaning of some terminology often used in supersymmetry.

The superpace is the generalization of usual spacetime in order to accommodate Grassmann variables.  Thus, we start by fixing our conventions for the operations with Grassmann variables. For a set of Grassmann variables, $\{ \theta_i,\theta_j\}=0$, we define
\begin{equation}
 \frac{\partial}{\partial \theta_i} (1)\equiv\int d\theta_i (1)\equiv0~~~\text{and}~~~\int d\theta_i \theta_j\equiv \frac{\partial}{\partial \theta_i} \theta_j\equiv\delta_{ij},
 \label{a1.1}
\end{equation}
and 
\begin{equation}
\frac{\partial}{\partial \theta_i} (\theta_j \theta_k)\equiv\left( \frac{\partial}{\partial \theta_i} \theta_j\right)\theta_k - \theta_j\left(\frac{\partial}{\partial \theta_i}\theta_k\right)= \delta_{ij}\theta_k-\delta_{ik}\theta_j.
\label{a1.2}
\end{equation}
 
Our goal here is to discuss the superpace formulation for a quantum mechanical system involving a single bosonic degree of freedom, $x(t)$. This is a (0+1)-dimensional theory whose ``spacetime" corresponds only to the time, $t$. The simplest superspace consists, in addition to the time coordinate, of a real Grassmann variable, $\theta$. This is called the $\mathcal{N}=1$ superspace.

As we shall see, the supersymmetry transformations correspond to translations in the superspace, 
\begin{equation}
t\rightarrow t'=t+i\epsilon\theta ~~~\text{and}~~~\theta\rightarrow \theta^{\prime}=\theta+\epsilon,
\label{a1.3}
\end{equation}
where $\epsilon$ is the infinitesimal Grassmannian parameter of the translation, which are generated by the supercharge
\begin{equation}
Q=\frac{\partial}{\partial\theta}+i\theta\frac{\partial}{\partial t}.
\label{a1.4}
\end{equation}

Now we can consider a real scalar function in the superspace, $\Phi(t,\theta)$, which is called a real scalar superfield. As $\theta^2=0$, its expansion in powers of $\theta$ is quite simple,
\begin{equation}
\Phi(t,\theta)=x(t)+i\theta\psi(t),
\label{a1.5}
\end{equation}
where $x(t)$ is identified as the usual bosonic degree of freedom and $\psi(t)$ is a dynamical fermionic degree of freedom. We see that the superfield simultaneously incorporates the bosonic and fermionic degrees of freedom. Furthermore, the number of bosonic and fermionic degrees of freedom is the same in the superfield, which is a basic requirement of the supersymmetry. We shall go back to this point in the case of extended supersymmetry.
Being a scalar, under the translations in (\ref{a1.3}), the superfield should transform as $\Phi'(t',\theta')=\Phi(t,\theta)$. By computing the left hand side, 
\begin{equation}
\Phi'(t+i\epsilon\theta,\theta+\epsilon)=\Phi'(t,\theta)+ i\epsilon\theta \frac{\partial}{\partial t}\Phi'+\epsilon \frac{\partial}{\partial \theta} \Phi',
\label{a1.6}
\end{equation}
we see that there is an arbitrariness concerning the last term, as we could have written it with the parameter $\epsilon$ to the right of the derivative, i.e., $\frac{\partial}{\partial \theta} \Phi'\epsilon$. As $\{\frac{\partial}{\partial\theta},\epsilon \}=0$, this will produce a minus sign if we try to let it in the above form. In this work, we always use the definition as in (\ref{a1.6}) for the Taylor expansion involving anticommuting variables. 

According to (\ref{a1.6}), it follows that 
\begin{equation}
\delta_{\epsilon}\Phi\equiv \Phi'(t,\theta)- \Phi(t,\theta)=-\epsilon Q \Phi. 
\label{a1.7}
\end{equation}
By comparing both sides, we see that the transformation of the components $x$ and $\psi$ are
\begin{equation}
\delta_{\epsilon}x=-i\epsilon \psi~~~\text{and}~~~\delta_{\epsilon}\psi=\epsilon \dot{x}.
\label{a1.8}
\end{equation}
These are the supersymmetry transformations. They in general take a boson into a fermion and a fermion into a boson. This is the sense of a symmetry between bosons and fermions. 

It is interesting to observe that supersymmetry ties internal and geometric (spacetime) symmetries, 
\begin{equation}
\{Q,Q\}=2i\frac{\partial}{\partial t}.
\label{a1.9}
\end{equation}
In other words, the anticommutator of the supercharges is proportional to the generator of the time translations, i.e., the Hamiltonian of the system.

There is another operator that plays an important role in the construction of an action in the superspace, called the supercovariant derivative,
\begin{equation}
D=\frac{\partial}{\partial\theta}-i\theta\frac{\partial}{\partial t}.
\label{a1.10}
\end{equation}
Note that it satisfies the following relation $\{Q,D\}=0$. This implies that $D\Phi$ transforms under supersymmetry as the superfields itself, $\delta_{\epsilon}(D\Phi)=-\epsilon Q (D\Phi)$. The same is true for the time derivative of the superfield, $\dot{\Phi}$, i.e., $\delta_{\epsilon} \dot{\Phi}=-\epsilon Q\dot{\Phi}$. Thus, any action written in terms of $\Phi$, $D\Phi$ and $\dot{\Phi}$ will be manifestly supersymmetric. Indeed, consider the action in the superspace,
\begin{equation}
S=\int dt d\theta \mathcal{L}(\Phi,D\Phi,\dot{\Phi}). 
\label{a1.11}
\end{equation}
We note here that the Lagrangian, $\mathcal{L}$, should be itself a Grassmann number in order to produce a nonvanishing scalar action.  Under a supersymmetry transformation, the variation automatically vanishes,
\begin{equation}
\delta_{\epsilon}S=\int dt d\theta \delta_{\epsilon}\mathcal{L}=\int dt d\theta (-\epsilon Q)\mathcal{L}=0,
\label{a1.12}
\end{equation}
since it is a total derivative. 

A simple action we can propose is
\begin{equation}
S= \frac{i}{2}\int dt d\theta  \dot{\Phi} D\Phi=\int dt \left(\frac{1}{2}\dot{x}^2-\frac{i}{2}\dot{\psi} \psi \right),
\label{a1.13}
\end{equation}
which is the action of the supersymmetric free particle. It is easy to check that it is invariant under the transformations in  (\ref{a1.8}). Now we could attempt to construct potential-like terms, for example, a term proportional to $x^2$. This is important for our purposes since the interaction part of the spherical model, $J_{{\bf r},{\bf r}'}S_{\bf r} S_{{\bf r}'}$, is essentially of this form. The conclusion is that with $\mathcal{N}=1$ we cannot construct such a term and we have to consider extended supersymmetry.

We then move to the case of two supersymmetries, $\mathcal{N}=2$. In this case, the superspace is higher dimensional and contains, in addition to the time, two Grassmann variables, $\theta$ and $\bar{\theta}$, which can be thought as the complex conjugate of each other. So, we can consider two independent translations in the superspace\footnote{Actually, it is the combination of these two transformations that produces a real translation for the time coordinate $t\rightarrow t +i(\bar\theta\epsilon-\bar\epsilon\theta)$, but we can consider one at a time.}, which will correspond to two independent supersymmetry transformations,
\begin{eqnarray}
&&\epsilon:~~~t\rightarrow t+i\bar\theta\epsilon,~~~\theta\rightarrow\theta-\epsilon,~~~\text{and}~~~\bar{\theta}\rightarrow\bar{\theta}\nonumber\\
&&\bar{\epsilon}:~~~t\rightarrow t-i\bar\epsilon\theta,~~~\theta\rightarrow \theta,~~~\text{and}~~~\bar\theta\rightarrow\bar\theta-\bar\epsilon,
\label{a1.14}
\end{eqnarray}
where $\epsilon$ and $\bar{\epsilon}$ are infinitesimal Grassmannian parameters. These translations are generated by the two supercharges
\begin{equation}
Q\equiv -\frac{\partial}{\partial\bar\theta}-i\,\theta\frac{\partial}{\partial t}~~~
\text{and}~~~\bar{Q}\equiv \frac{\partial}{\partial\theta}+i\,\bar\theta\frac{\partial}{\partial t},
\label{a1.15}
\end{equation}
satisfying the following anticommutation relations
\begin{equation}
\{Q,Q\}=0,~~~\{\bar{Q},\bar{Q}\}=0~~~\text{and}~~~\{Q,\bar{Q}\}=-2i\frac{\partial}{\partial t}.
\label{a1.16}
\end{equation}

Now we consider a real scalar superfield, $\Phi(t,\theta,\bar\theta)$. Its expansion in powers of the Grassmann variables has more components than the previous case, 
\begin{equation}
\Phi(t,\theta,\bar{\theta})=x+\bar{\theta}\psi+\bar{\psi}\theta+\bar{\theta}\theta F.
\label{a1.17}
\end{equation} 
We have two bosonic degrees of freedom, $x$ and $F$, and two fermionic degrees of freedom, $\psi$ and $\bar{\psi}$. Notice again the matching of the fermionic and bosonic degrees of freedom. However, in contrast to the case $\mathcal{N}=1$, an important distinction concerning the matching of degrees of freedom takes place now. To appreciate this, notice that so far there is nothing about equations of motion, which dictate the dynamics of the model. Thus, we say that the matching is off-shell. Not necessarily all the variables present in the superfield correspond to physical degrees of freedom.  For a simple action that we will consider, the resulting equations of motion imply that $F$ is an auxiliary degree of freedom whereas the fermionic degrees of freedom $\psi$ and $\bar{\psi}$ are not independent. We shall discuss in a moment.

Under translations, the scalar superfield transforms as $\Phi'(t',\theta',\bar{\theta}')=\Phi(t,\theta,\bar{\theta})$. By using the same definition for the Taylor expansion as in (\ref{a1.6}), we obtain for the functional variations
\begin{equation}
\delta_{\epsilon}\Phi=-\bar{Q}\epsilon\Phi~~~\text{and}~~~\delta_{\bar{\epsilon}}\Phi=-\bar{\epsilon}Q\Phi.
\label{a1.18}
\end{equation}
These relations lead the following transformations for the components
\begin{equation}
\epsilon:~~~\delta_{\epsilon}x=\bar{\psi}\epsilon,~~~
\delta_{\epsilon}\psi=-i\dot{x}\epsilon+
F\epsilon,~~~
\delta_{\epsilon}\bar{\psi}=0,~~~\text{and}~~~\delta_{\epsilon}F=i\dot{\bar\psi}\epsilon;
\label{a1.19}
\end{equation}
and
\begin{equation}
\bar{\epsilon}:~~~\delta_{\bar\epsilon}x=\bar\epsilon{\psi},~~~
\delta_{\bar\epsilon}\psi=0,~~~
\delta_{\bar\epsilon}\bar{\psi}=i\dot{x}\bar\epsilon+
F\bar\epsilon,~~~\text{and}~~~\delta_{\epsilon}F=-i\bar\epsilon\dot{\psi}.
\label{a1.20}
\end{equation}
The supercovariant derivatives are constructed by taking the opposite combinations of derivatives of the supercharges in (\ref{a1.15}), 
\begin{equation}
D\equiv -\frac{\partial}{\partial\bar\theta}+i\theta\frac{\partial}{\partial t}~~~
\text{and}~~~\bar{D}\equiv \frac{\partial}{\partial\theta}-i\bar\theta\frac{\partial}{\partial t},
\label{a1.21}
\end{equation}
satisfying the following anticommutation relations with the supercharges,
\begin{equation}
\{D,Q\}=\{D,\bar{Q}\}=\{\bar{D},Q\}=\{\bar{D},\bar{Q}\}=0.
\label{a1.22}
\end{equation}
As already discussed in the case $\mathcal{N}=1$, these relations ensure that any action in the superspace involving superfields and supercovariant derivative of superfields is manifestly supersymmetric. In general, we have
\begin{equation}
S=\int dt d\theta d\bar{\theta} \mathcal{L}(\Phi, \dot{\Phi},D\Phi,\bar{D}\Phi),
\label{a1.23}
\end{equation}
where, in contrast to the case $\mathcal{N}=1$, the Lagrangian is a scalar. 

We can propose a simple form for the action,
\begin{equation}
S=\int dt d\theta d\bar{\theta} \left(\frac{1}{2} \bar{D}\Phi  D\Phi - U(\Phi)\right),
\label{a1.24}
\end{equation}
where $U$ is an arbitrary function of the superfield. To write the action in term of the components,  we expand the potential $U$ in powers of $\theta$ and $\bar{\theta}$ and then select the $\bar{\theta}\theta$ contribution, 
\begin{equation}
U(x+\bar{\theta}\psi+\bar{\psi}\theta+\bar{\theta}\theta F)\Big{|}_{\bar{\theta}\theta}=U'(x) F - U''(x) \bar{\psi}\psi,
\label{a1.25}  
\end{equation}
where the primes mean derivatives with respect to $x$. Thus, in components, the action reads
\begin{equation}
S=\int dt \left(\frac{1}{2}\dot{x}^2+i\bar{\psi}\dot{\psi}+\frac{1}{2}F^2- U'(x) F + U''(x) \bar{\psi}\psi \right).
\label{a1.26}
\end{equation}
This is the off-shell action for the supersymmetric quantum mechanics with $\mathcal{N}=2$. It is easy to verify that it is invariant under (\ref{a1.19}) and (\ref{a1.20}). 
This expression shows that $F$ is an auxiliary degree of freedom, as there is no time derivative of $F$. Its equation of motion is just an algebraic one,
\begin{equation}
F=U',
\label{a1.27}
\end{equation}
and can be eliminated from the Lagrangian, leading to the on-shell formulation,
\begin{equation}
S=\int dt \left(\frac{1}{2}\dot{x}^2+i\bar{\psi}\dot{\psi}-\frac{1}{2}( U'(x))^2 + U''(x) \bar{\psi}\psi \right).
\label{a1.28}
\end{equation}
In this form it is immediate to construct potential terms of interest. For example, the harmonic potential corresponds to $U'=\omega x$, which, in turn, comes from the following term in the superspace:
\begin{equation}
U(\Phi)=\frac{1}{2}\omega \Phi^2.
\label{a1.29}
\end{equation}

As a last point, we discuss the counting of degrees of freedom in the on-shell formulation. At first sight, it seems that there is no matching, since we have one bosonic degree of freedom, $x$, and two fermionic degrees of freedom, $\psi$ and $\bar{\psi}$. However, their equations of motion are,
\begin{equation}
i\dot{\bar{\psi}}+U''\bar{\psi}=0~~~\text{and}~~~i\dot{\psi}-U''\psi=0.
\label{a1.30}
\end{equation}
They are just the complex conjugated of each other. By writing $\psi=\psi_1+i \psi_2$, with $\psi_1$ and $\psi_2$ being real Grassmann variables, we see that they are not independent at all, 
\begin{equation}
\dot{\psi}_2+U''\psi_1=0~~~\text{and}~~~\dot{\psi}_1-U''\psi_2=0.
\label{a1.31}
\end{equation} 
Thus we see that in fact we have only one independent fermionic degree of freedom. The key point is that the matching of degrees of freedom is achieved only after the use of the equations of motion in the on-shell formulation. This is a general property of supersymmetric theories. We note also that there is no such a distinction in the case with $\mathcal{N}=1$ discussed previously. 

To summarize the case $\mathcal{N}=2$, the superspace formalism naturally delivers the off-shell formulation, where the matching of the degrees of freedom is automatic, at the price of the introduction of auxiliary variables. On the other hand, the on-shell formulation deals with physical degrees of freedom and the matching is reached upon using of equations of motion.

%%%%%%%%%%%%%%%%%%%%%%%%%%%%%%%%%%%%%%%%%%%%%%%%%%%%%%%%%%%

%%%%%%%%%%%%%%%%%%%%%%%%%%%%%%%%%%%%%%%%%%%%%%5

%%%%%%%%%%%%%%%%%%%%%%%%%%%%%%%%%%%%%%%%%%%%%%%%%%%%%%%%%%%%%%%%%%%%%%%%%%%%%%%%%%%%%%%%%%%%%%%%%%%%%%%%%%%%%%%%%5
\end{document}